\documentclass[12pt,reqno]{amsart}

\usepackage[utf8]{inputenc}
\usepackage[english]{babel}

\usepackage{hyphenat}
\usepackage{hyperref}

\usepackage{float}

\usepackage{graphicx}
\usepackage{xcolor}

\definecolor{gray}{rgb}{0.5,0.5,0.5}

\newtheorem{statement}{It can be verified that:}

\usepackage{amsfonts,amssymb,amsbsy,mathabx}
\usepackage{upgreek}
\usepackage{colonequals}

\usepackage{pifont}
\usepackage{yfonts}
\usepackage{textcomp}
\usepackage{txfonts}  

\newcommand{\half}{\mbox{\scriptsize$1\over2$}}
\newcommand{\Ratio}[2]{\mbox{\small$#1\over#2$}}
\newcommand{\Eq}{Eq.~}
\newcommand{\Eqs}{Eq.s~}
\newcommand{\Tab}{Tab.~}
\newcommand{\Src}{Ref.~}

\newcommand{\Pain}{Painlev\'e}
\newcommand{\PainIII}{Third Painlev\'e}
\newcommand{\PIIItran}{\hbox{{\sf P}$_{\mbox{\tiny III\,}}$-\,function}}
\newcommand{\PIIIeq}  {\hbox{{\sf P}$_\mathrm{III}\,$-\,equation}}

\newcommand{\Imi}{\mathrm{i}}

\newcommand{\eiphi}{{\Phi}}
\newcommand{\eP}{{\Psi}}

\newcommand{\bflam}{\pmb{\lambda}}
\newcommand{\bfmu}{\pmb{\mu}}

\newcommand{\ddbflam}{\hspace{0.2ex}\bflam\hspace{-1.4ex}\raisebox{0.8ex}{\textdegree\hspace{-0.2ex}\textdegree}\hspace{0.4ex}}
\newcommand{\dbflam}{\hspace{0.2ex}\bflam\hspace{-1.03ex}\raisebox{0.8ex}{\textdegree}\hspace{0.6ex}}

\newcommand{\dbfmu}{\hspace{0.0ex}\bfmu\hspace{-1.ex}\raisebox{0.3ex}{\textdegree}\hspace{0.3ex}}

\newcommand{\dbf}[1]{#1\hspace{-0.ex}\raisebox{0.3ex}{\textdegree}\hspace{0.3ex}}

\newcommand{\bfuplam}{\pmb{\uplambda}}
\newcommand{\bfupmu}{\pmb{\upmu}}

\newcommand{\dbfuplam}{\hspace{0.2ex}\bfuplam\hspace{-1.3ex}\raisebox{0.9ex}{\textdegree}\hspace{0.5ex}}
\newcommand{\ddbfuplam}{\hspace{0.2ex}\bfuplam\hspace{-1.5ex}\raisebox{0.85ex}{\textdegree\hspace{-0.15ex}\textdegree}\hspace{0.2ex}}
\newcommand{\dbfupmu}{\hspace{0.0ex}\bfupmu\hspace{-1.2ex}\raisebox{0.3ex}{\textdegree}\hspace{0.4ex}}
\newcommand{\ddbfupmu}{\hspace{0.0ex}\bfupmu\hspace{-1.55ex}\raisebox{0.4ex}{\textdegree\hspace{-0.15ex}\textdegree}\hspace{0.3ex}}

\newcommand{\dualam}{\reflectbox{$\pmb{\uplambda}$}}
\newcommand{\duamu}{\reflectbox{$\pmb{\upmu}$}}

\newcommand{\dddualam}{\hspace{0.2ex}\dualam\hspace{-1.2ex}\raisebox{0.85ex}{\textdegree\hspace{-0.2ex}\textdegree}\hspace{0.2ex}}
\newcommand{\ddualam}{\hspace{0.2ex}\dualam\hspace{-0.8ex}\raisebox{0.85ex}{\textdegree}\hspace{0.3ex}}
\newcommand{\ddduamu}{\hspace{0.2ex}\duamu\hspace{-1.35ex}\raisebox{0.4ex}{\textdegree\hspace{-0.2ex}\textdegree}\hspace{0.3ex}}
\newcommand{\dduamu}{\hspace{0.2ex}\duamu\hspace{-1.07ex}\raisebox{0.25ex}{\textdegree}\hspace{0.5ex}}

\newcommand{\dakav}{\hspace{0.2ex}\mbox{\tt v}\hspace{-1.10ex}\raisebox{0.25ex}{\textdegree}\hspace{0.5ex}}
\newcommand{\ddv}{\hspace{0.2ex}\mbox{\tt v}\hspace{-1.35ex}\raisebox{0.4ex}{\textdegree\hspace{-0.2ex}\textdegree}\hspace{0.3ex}}
\newcommand{\akav}{\mbox{\tt{v}}}

\newcommand{\rmG}{\mathrm{G}}
\newcommand{\drmG}{ \mathrm{G}^{\!\scalebox{0.7}{\rotatebox{18}{$\mathtt{\prime}$}}} }
\newcommand{\ddrmG}{\mathrm{G}^{\!\scalebox{0.7}{\rotatebox{18}{$\mathtt{\prime}$}}\!\scalebox{0.7}{\rotatebox{18}{$\mathtt{\prime}$}} } }

\newcommand{\drmchar}[1]{{#1}^{\!\scalebox{0.7}{\rotatebox{18}{$\mathtt{\prime}$}}} }

\newcommand{\akaz}{\mbox{\tt{z}}}
\newcommand{\akat}{\mbox{\tt{t}}}

\newcommand{\troot}{\mbox{\mbox{\tiny$\mathstrut_0$}\!\akat}}
\newcommand{\tpole}{\mbox{\mbox{\tiny$\mathstrut_\infty$}\!\akat}}

\newcommand{\akatau}{{{\tau}}}

\newcommand{\ddbfG}{\mathbf{G}^{\!\scalebox{0.7}{\rotatebox{18}{$\mathtt{\prime}$}}\!\scalebox{0.7}{\rotatebox{18}{$\mathtt{\prime}$}} } }
\newcommand{\dbfG}{\mathbf{G}^{\!\scalebox{0.7}{\rotatebox{18}{$\mathtt{\prime}$}} } }
\newcommand{\bfG}{\mathbf{G}}

\newcommand{\llll}{{\ell}}
\newcommand{\halfi}{\mbox{\scriptsize$\Imi\over2$}}

\newcommand{\modtwopi}{\mbox{${mod}\:2\pi$}}

\newcommand{\rotnum}{\mbox{{\textswab n}}}

\newcommand{\asteRisk}{\bullet}

\newcommand{\Tau}{\mathcal T}

\newcommand{\Fig}{{Fig.~}}

\begin{document}


\author{ S.I.~Tertychniy
\vspace{-3em}
}

\title[\sc transportation of constrictions]%
{\sc
On transportation of constrictions on phase-lock areas
of overdamped Josephson junction
by means of
isomonodromic deformation
of the associated special double confluent Heun equation
}

\maketitle

\begin{center}{\sc vniiftri}
\end{center}

\begin{abstract}
An  example of computation of locations of constrictions  on the phase-lock area in the parameter space for the equation modeling dynamics of overdamped Josephson junction is discussed. Several  conjectures of qualitative character are inferred from the heuristic data thus obtained. The method applying isomonodromic deformations of the special double confluent Heun equation associated with the noted model is used. Such a deformation is controlled by the data provided by solution to a Third \Pain{} equation for which a new representation in conjunction with the corresponding non-linear second order differential equation fairly efficient in the treatment of the problem in question is proposed.
\end{abstract}

\section{Preliminaries:}
\subsection{Background and formulas}

\hphantom{ }\\[1ex]
Notwithstanding the development of various designs
and usage of different technologies,
Josephson junction
can be thought of as
an electronic component
constituted by three small pairwise contiguous layers of which the two outer ones
consist of superconducting material while the thinner layer inbetween them is of a distinct nature
and can consist of an insulator, of a ``normal'', non-superconducting metal,  
or, sometimes, to be of a more compounded structure.
The superconducting layers are set from outside in contact with some external circuit through which
a varying electric current of certain profile is brought.
Another way of characterization of the junction state is the indication of the electric
voltage arising across or applied to it by an external equipment.

In many cases including ones of practical interest the behavior of Josephson junction
under external influence
can be modeled by
the following non-linear differential equation 
 \begin{equation} \label{RSJ or RCSJ}%
{\hbar C\over 2 e}
{d^2\over d t^2}\varphi(t)
+
{\hbar G\over 2 e}
  {d\over d t} \varphi(t)
+
{\mathcal I}_c
\sin\varphi(t) = I(t) 
\end{equation}
(see, e.g., \Src\cite[\Eq{}(10.24)]{MK}, where distinct notation of the unknown function is used). 
The following universal physical constants are here involved in the summand coefficients:
$\hbar$ is the reduced Planck constant, $e$ is the electron charge.
Other constant parameters depend on junction properties. Of them,
 $C$ and  $G$ stand for the effective values of junction capacitance and conductance, respectively.
The constant ${\mathcal I}_c$ denotes the junction critical current, i.e.\ the maximal direct current that
is realized by the quantum tunneling of Cooper pairs of electrons in superconducting layers 
(supercurrent)
through the non-superconducting one separating them
in case of zero
difference of the electric potential.

The above equation binds the two function of time $t$.
Of them, $I(t)$, represents the intensity of electric current through the junction which is often called bias current.
The function $\varphi(t)$ is called “the difference of phases of the order
parameter” or, more concisely, "the phase difference", where "the order parameter” has the physical meaning of the collective
(macroscopic) wave function for the Cooper pairs of electrons present in superconducting layers of junction.

The observable manifestation of evolution in time of the phase difference is embodied in the equation 
\begin{equation}  \label{momentaryV}
 V(t)
%
 = {\Phi_0\over 2\pi}{d\over d t} \varphi(t).
\end{equation}
Here $V(t)$ denotes the voltage applied to junction or observed as the difference of electric potentials on its opposite outer layers.
$\Phi_0 = (2e)^{-1} h $ is the another universal physical constant named the magnetic flux quantum ($h=2\pi\hbar$ is the `original' Plank constant).

Equation \eqref{RSJ or RCSJ} plays role of the mathematical base of the so called RSJ model of a Josephson junction (Resistively Shunted Junction);
in more modern texts the acronym RCSJ (Resistively and Capacitively Shunted Junction, respectively) can also be found.
See Refs
\cite[Eq. (2.2)]{McC}
\cite[Eq.\ (1)]{Ste}
\cite[Eq.s\ (1),(2)]{Ka}
\cite[Sec.\ 2.3, Eq.\ (2.24a)]{Li}
\cite[Sec.s 1.4,11,3, Eq. (11.3.1)]{BP}
\cite[Sec. 4.3,Eqs. (7.30)-(7.32)]{ShV} 
\cite[section 10.4.2]{MK}

We will consider here a simplistic version of \Eq\eqref{RSJ or RCSJ} which arises after removal
of the first summand in on the left, i.e.\ upon setting $C=0$. In physical terms, such a case corresponds
to conditions under which the influence of junction capacitance becomes negligible.
Josephson junctions meeting it are named overdamped.
In particular, the model of overdamped junction is usually used in case of so called SNS  ({\small Superconductive-Normal-Superconductive metal layers})
Josephson junctions \cite{SMWKetal}.

It is also required of the function $I(t)$ to be periodic and representable as a Fourier series with only two initial terms present.
Then, without loss of generality, 
\Eq\eqref{RSJ or RCSJ} simplified in this way can be recast to the equation
\begin{equation}
\label{overdamped RSJ}
 \dot \phi(\akatau) + \sin\phi(\akatau) = B + A\cos\omega \akatau,
\end{equation}
\begin{equation} \label{defs of tau et al}
\hspace{-1em}
\begin{aligned}
\mbox{where }&
\akatau
  =\frac{2\pi {\mathcal I}_c} {\Phi_0 G}\, t,\;
\phi(\akatau) =\varphi(t), \;
\dot{\vphantom{I}{\color{gray}\square}} ={d{ {\color{gray}\square} } \over d \akatau},\;  \hfill
(B + A\cos\omega \akatau)\,{\mathcal I}_c = I(t)
\end{aligned}\hspace{-1em}
\end{equation}
The new constants $A,B,\omega$ introduced above will be considered real.
They can be interpreted as the dimensionless measures of the amplitude (hence $A$) of injected AC
and the constant bias (hence $B$) by additional DC, both given in units of the critical current $ {\mathcal I}_c $.
Dimensionless $\omega$ of which is required to be positive
is the frequency of the noted AC with respect to the rescaled time $\akatau$. 

\Eq\eqref{overdamped RSJ} is essentially equivalent to the  Riccati equation
\begin{eqnarray}
          \label{Riccati}
&&
z \,{\eiphi}'
+(2\Imi \omega)^{-1}({\eiphi}^2-1)=
\big(\llll+\mu\,(z+z^{-1})\big){\eiphi}
\end{eqnarray}
constraining holomorphic function $ \eiphi= \eiphi(z)$,
where and also in what follows $z$ denotes a free complex variable and
${\color{gray}\square}' = d{\color{gray}\square}  / d z$.
The constant parameters $\llll,\mu$  are in 1-to-1 correspondence to the constants characterizing \Eq\eqref{overdamped RSJ}
in accordance with the
equalities
\begin{equation}
 \label{AB vs lamu}
A=2\omega\,\mu, \; B=\omega\,\llll
\end{equation}

In vicinity of the point $z=1$ \Eq\eqref{overdamped RSJ} arises from \Eq\eqref{Riccati} as the result of  restriction of the latter to
a segment of unit circle containing $1$ in accordance with the equality
\begin{equation} \label{from circle to C}
e^{\Imi\phi(\akatau)}
= \eiphi(z)
\mbox{ taking place for } z=e^{\Imi\omega \akatau},
\end{equation}
where the real variable $\akatau$ is assumed to vary in vicinity of zero.
The resulting function $\phi(\akatau)$
is real valued, provided $|\eiphi(1)|=1.$

Conversely, the function $ \eiphi(z) $    arises a the result of analytic continuation of the function
taking values
$e^{\Imi\phi(\akatau)}  $ at points $z=e^{\Imi\omega \akatau}$ for real $\akatau$ varying in vicinity of zero.
 It can also be obtained as the local solution of the Cauchy problem for \Eq\eqref{Riccati} with initial value $\eiphi(1)= e^{\Imi\phi(0)}$.

The linear differential equation
\begin{equation}
  \label{eq for e^P}
2\Imi\,\omega\, z\,\eP'=( {\eiphi} + {\eiphi}^{-1})\eP.
\end{equation}
in conjunction 
with the constraint (the initial data)
$\eP(1)=1$
uniquely define 
the non-zero
holomorphic function $ \eP= \eP(z)$, provided holomorphic function $ \eiphi(z)$ has been determined.
As a consequence of
its restriction
to the unit circle where  $z=e^{\Imi\omega \akatau} $, the real-valued function
\begin{equation}
P(\akatau) = \log \eP(e^{\Imi\omega \akatau })
\end{equation}
obeys the equation $ \dot P=\cos\phi$ implying that
\begin{equation}\label{P-integral}
 P(\akatau) = \int_0^\akatau \cos\phi(\tilde \akatau) d\tilde \akatau
\end{equation}
In addition, given the complex-valued function $\mathcal{F}=Q(\akatau)+\Imi e^{-P(\akatau)}$ with real $Q$ obeying the equation
$
\dot{\mathcal{F}}= - \Imi e^{\Imi\phi} \,\mathrm{Im}\, \mathcal{F} \:
$
with initial condition $ \mathcal{F}(1)=\Imi$,
one can obtain that 
$ \dot Q =  e^{-P} \sin \phi $
leading to the explicit representation
\begin{equation}\label{Q-integral}
 Q(\akatau) = \int_0^\akatau e^{-P(\tilde \akatau)}\sin\phi(\tilde \akatau) d\tilde \akatau
\end{equation}

The above definitions enable one to introduce
the following expression (non-linear functional)
\begin{equation}\label{mprediscr}
\begin{aligned}[b]
\Lambda=\Lambda[\phi] =\Lambda(A,B,\omega) =&\;
\cosh\half\big( P(T)-P(0)\big)\cos\half\big(\phi(T)-\phi(0)\big)\\
&\;
\hspace{-6.6ex}\hspace{-5.5ex}
- \half e^{\left( P(T)+P(0)\right)/2}\big(Q(T)-Q(0)\big) \sin\half\big(\phi(T)-\phi(0)\big)
\end{aligned}\
\hspace{-3em}
\end{equation}
which
is used for determination of the sort of asymptotic behavior of solutions to \Eq\eqref{overdamped RSJ}.
The integrals defining $P$ and $Q$ by the formulas
\eqref{P-integral}, \eqref{Q-integral} may also be regarded as indefinite ones.
The function $\phi(\akatau)$ may be arbitrary real solution to the mentioned equation.
The value of $\Lambda$ does not depend on the particular choice of the latter.

The role of the value of the expression \eqref{mprediscr} in prediction of propertied of solutions to \Eq\eqref{overdamped RSJ} is as follows.
If $\Lambda >1$ then there are such solutions $\phi(\akatau)=\phi_{(\pm)}(\akatau)  $ to \Eq\eqref{overdamped RSJ} that
the exponents $ e^{\Imi\phi}$ are periodic with period $T=2\pi\omega^{-1}$.
 They can be expressed
via solution $\phi_0(\akatau)$ to  \Eq\eqref{overdamped RSJ} with zero initial data, $\phi_0(0)=0$,
by the formula
\begin{equation}
e^{\Imi\phi(\akatau)}=e^{\Imi\phi_0(\akatau)} {1 + c\cdot(Q_0(\akatau) - \Imi e^{-P_0(\akatau)})\over 1 + c\cdot(Q_0(\akatau) + \Imi e^{-P_0(\akatau)})  }
\end{equation}
where $P_0,Q_0$ are defined
by formulas \eqref{P-integral}, \eqref{Q-integral} upon formal replacements
$\phi\leftleftharpoons\phi_0, P\leftleftharpoons P_0, Q\leftleftharpoons Q_0  $.
The constant $c$ has to take
one of the two values
\begin{equation}\label{periodicity}
\hspace{-2em}
 \begin{aligned}[t]
c_{(\pm)}=&\;
   e^{-P_0(T) /2}\sin\half\phi_0(T)\times
\\
&\hspace{-2em}
\Big(
-\sinh\half P_0(T)\cos\half\phi_0(T) +\half e^{P_0(T)/2}Q_0(T)\sin\half\phi_0(T)
\pm\sqrt{\Lambda^2-1}\Big)^{-1}\rlap{}\hspace{-4em}
\end{aligned}
\end{equation}
of which one corresponds to a stable \mbox{${mod}\:2\pi$}-periodic solution to \Eq\eqref{overdamped RSJ}
while another provides a similar but unstable one.
Besides, it holds
\begin{equation}\label{increment}
 \begin{aligned}[t]
\phi(T)-\phi(0)
=
\phi_0(T)
+
2\, c_{}\cdot \mbox{\sf Re}
\int_0^T
{  e^{-P_0(\akatau)}   e^{-\Imi\phi_0(\akatau)}    d \,\tau
 \over
1 + c_{} \cdot (Q_0(\akatau) - \Imi e^{-P_0(\akatau)} )
}
\end{aligned}
\end{equation}
This difference is integer-proportional to $2\pi$ and its values in cases $c=c_{(+)}$ and $c=c_{(-)}$ coincide.

\bigskip

Let us further construct
holomorphic functions
$E_{\mbox{\tiny$\{\!+\!\}$}}, E_{\mbox{\tiny$\{\!-\!\}$}}$
 in accordance with the following dual definition
\begin{equation}
              \label{sharpeigens}
\begin{aligned}
E_{\mbox{\tiny$\{\!\pm\!\}$}}\!(z)&
\colonequals 
2^{-1}
e^{\mu(z+z^{-1}-2)/2}
z^{-\llll/2}
\times\\&\hspace{2ex}
\left\lgroup
\frac{1\pm\Imi}{\sqrt2}
\big(\eP(z)\eiphi(z)\big)^{1/2}
+
\frac{1\mp\Imi}{\sqrt2}
\big(\eP(z^{-1})/\eiphi(z^{-1})\big)^{1/2}
\right\rgroup
\end{aligned}
\end{equation}
The above expressions are obviously well defined in vicinity of $z=1$.
Since they imply that 
$E_{\mbox{\tiny$\{\!\pm\!\}$}}(1)=\cos\half(\phi(0)\pm\pi/2)$
the functions 
$E_{\mbox{\tiny$\{\!\pm\!\}$}}  $
are not identically null as long as $\phi(0)\not=\pi/2\; (\mbox{mod }\pi)$.
Moreover, it is easy to see that if
$E_{\mbox{\tiny$\{\!\pm\!\}$}}(1)=0$ then
$E_{\mbox{\tiny$\{\!\mp\!\}$}}(1)$ is equal  to either $1$ or $-1$.
Accordingly, they cannot both concurrently identically vanish.


On condition of fulfillment of \Eqs{ }\eqref{Riccati}, \eqref{eq for e^P},
the functions $ E_{\mbox{\tiny$\{\!+\!\}$}} $, $ E_{\mbox{\tiny$\{\!-\!\}$}} $
prove to be
the eigenvectors
of the linear operator $ \varhash $ defined as follows.
\begin{equation}\label{diez}
  \varhash\!: E=E(z) \to \varhash E =(\varhash E)(z)=2\omega\, z^{-\llll-1} \big(E'(z^{-1})-\mu E(z^{-1})\big)
\end{equation}
The corresponding eigenvalues are equal to $+1$ and $-1$, respectively.
This means, in particular, that
non-identically-zero functions $E_{\mbox{\tiny$\{\!+\!\}$}}$ and $E_{\mbox{\tiny$\{\!-\!\}$}}$  are linearly independent.

On the other hand,
it is easy to show \cite{T1}
that such eigenfunctions of the map $\varhash$ have to obey the following linear homogeneous second order differential equation
\begin{equation}
\label{sDCHE} %
0=
z^2 E''
+\big((\ell+1)\,
z
+  (1-z^2)\,\mu
\big) E'
+\big
( -(\ell+1)\,\mu\, z
+
\lambda
\big
)
E
\end{equation}
in which the constant $\lambda$ is determined by the constraint
\begin{equation} \label{constconstr}
4\omega^2(\lambda + \mu^2)=1\mbox{ implying }\lambda=(2\omega)^{-2} - \mu^2
\end{equation}
and which is a particular case of double confluent Heun equation \cite{SW}, \cite{SLS}.
In addition, the linear space of solution to \Eq\eqref{sDCHE} is preserved under the action of the operator $\varhash$
which represents its linear automorphism.

One sees therefore that any solution to  \Eq\eqref{overdamped RSJ} gives rise to some solutions to \Eq\eqref{sDCHE}
which are produced
by means of an explicit transformation
involving analytic continuation.

Conversely, let a non-trivial solution $E=E(z)$ to \Eq\eqref{sDCHE} with real parameters $\llll,\lambda, \mu$  which is real-valued on the real axis be given.
Then in vicinity of the point $z=1$
the following
 holomorphic function
\begin{eqnarray}
 \eiphi(z)
= - z^\llll
{
e^{\halfi\alpha}
E(z)
+
\Imi\, e^{-\halfi\alpha}(\varhash E)(z)
\over
e^{-\halfi\alpha}
E(z^{-1})
-
\Imi\, e^{\halfi\alpha}(\varhash E)(z^{-1})
},
\end{eqnarray}
where $\alpha$ is an arbitrary real constant,
verifies \Eq\eqref{Riccati}.
Moreover, 
then
$ |\eiphi(1)|=1 $ and
the transition \eqref{from circle to C}
defines, up to addition of an arbitrary integral number of $2\pi$,
the real-valued function $\phi(\akatau)$ obeying \Eq\eqref{overdamped RSJ}.

Additionally, the real-valued functions $P(\akatau),Q(\akatau)$
determined by
the restriction of the holomorphic function
%
\begin{eqnarray}
 \Theta(z)
=-\Imi
{
e^{-\halfi\alpha}E(z^{-1})
+
\Imi\, e^{\halfi\alpha}(\varhash E)(z^{-1})
\over
e^{-\halfi\alpha}
E(z^{-1})
-
\Imi\, e^{\halfi\alpha}(\varhash E)(z^{-1})
},
\end{eqnarray}
to the unit circle with subsequent splitting  according to the template 
\begin{equation} 
\Theta(e^{\Imi\omega\akatau}) = Q(\akatau) + \Imi e^{-P(\akatau)}
\end{equation}
coincides with the integrals \eqref{P-integral} and \eqref{Q-integral}, respectively.

The point $z=0$ is the only singularity for the equation \eqref{sDCHE} on the complex plane.
(When considered on the Riemann sphere, $z=\infty$ appears to be its singular point as well.)
Hence any its solution is holomorphic at any non-zero point of its domain.
On the contrary,
under condition of genericity, there are no functions obeying \Eq{\,}\eqref{sDCHE} and continuous at zero.
At the same time, for certain special choices of parameters its solutions holomorphic at zero and hence defining entire functions may exist \cite{T2}, \cite{BT1}.
One of such special conditions requires of the parameter $\ell$ to be integer. It is even possible
to consider only non-negative 
 since the replacement
$E(z)\leftleftharpoons (d/d\,z )^\ell E(a)$  of the unknown by its $\ell$-fold derivative
carries out only the inversion $\ell \leftleftharpoons  -\ell$
and maps therefore the set of entire non-polynomial solutions to \Eq{\,}\eqref{sDCHE}
to the set of entire non-polynomial solutions of the similar equation with opposite integer value of the parameter $\ell$ now negative.
A holomorphic solution to \Eq{\,}\eqref{sDCHE} with negative integer $\ell$, if any, is transformed
to holomorphic solution to similar equation with opposite (and thus positive integer) value of this parameter. 

In order to formulate another condition of existence of solution to \Eq{\,}\eqref{sDCHE} holomorphic at zero
(i.e, in the whole complex plane)
let us define  $2\times2$ matrices
\begin{equation}
                \label{M-matrix}
M_j= \begin{pmatrix}
1+\lambda/(j(j- l )) & \mu^2/ (j(j- l )) \\
1 & 0 \end{pmatrix}, 
\end{equation}
where the restriction $  l =0,1,2,\dots $ is assumed,
and the sequence of their infinite products
\begin{equation}
                   \label{R-matrix}
R_m=  \rlap{\color{gray}\mbox{\,$\lessdot$}}  \!\!\prod_{j=m}^\infty M_j \mbox{ with subscripts } m= l +1, l +2,\dots,
\end{equation}
which  converge for arbitrary $\lambda$  and $\mu$.
Here the mark `$\lessdot$' indicates implementation of indispensible relative arrangement of non-commuting matrix factors $ M_j $
ensuring ascending of their integer subscripts from left to right.

Then in case of non-negative integer $\ell$
the Taylor series
\begin{equation}
          \label{E-series}
\hphantom{.}\hspace{-13.2em}
E(z)
=\sum_{k=0}^{\infty} a_k z^k
\mbox{ with }
\end{equation}
\begin{eqnarray}
\vphantom{.}\nonumber
\\[-3ex]
a_0
&\!\!=\!\!&
\begin{pmatrix}1 & 0\end{pmatrix}
\cdot R_{\ell+1} \cdot
\begin{pmatrix}1 & 0\end{pmatrix}^{\!\!\intercal}%
\mbox{ and}%
\\
a_k
&\!\!=\!\!&
{\mu^k\over k!}
\begin{pmatrix}0 & 1\end{pmatrix}
\cdot R_{\ell+1 + k} \cdot
\begin{pmatrix}1 & 0\end{pmatrix}^{\!\!\intercal}
\mbox{ for }k=1,2,\dots
\end{eqnarray}
defines an entire function which verifies \Eq{\,}\eqref{sDCHE} if and only if
\begin{equation}\label{perfectness}
 \xi_\ell =0,
\end{equation}
 where
%
%
\begin{equation}
           \label{def of xi}
\hphantom{.}\hspace{-8.2em}
\xi_{\ell}(\lambda,\mu)
=
\begin{pmatrix}\lambda & \mu^2\end{pmatrix}
\cdot R_{\ell+1} \cdot
\begin{pmatrix}1 & 0 \end{pmatrix}^{\!\!\intercal}
\end{equation}
It is also useful to introduce its alternative implementation with modified arguments
defined as follows
\begin{equation}
               \label{withnotherarg}
\xi_{\ell}(\mu|\omega)=\xi_{\ell}((2\omega)^{-2}-\mu^2,\mu)
\end{equation}
\textit{ Cf\/} \Eq\eqref{constconstr}.

Under special conditions, \Eq\eqref{sDCHE} can also admit polynomial solutions \cite{T3}, \cite{BT2}.
They prove however to be not incorporated in
the set of holomorphic solutions \eqref{E-series}. Polynomial  solutions may only arise if $\llll$ is a negative integer
\begin{equation}
\llll= -1 - n, \; n \in \{0,\mathbb{N}\}
\end{equation}
Besides, the vanishing of the determinant $\Delta_n= \Delta_n(\lambda,\mu)$ of the tri-diagonal $(n+1)\times(n+1)$-matrix
 $\mathrm{V}_n = \mathrm{V}_n(\lambda,\mu) $ with elements (\cite{T3}, \Eq~(20))
%
%
\begin{equation}
\begin{aligned}
&v_{k,l} = 0 &&\mbox{ if }
|k-l|>1
\\ &
v_{k,k} =\lambda - k (n-k+1) &&\mbox{ for }k=0, \dots, n \mbox{ (the diagonal)}
\\ &
v_{k,k-1} = \mu\, k &&\mbox{ for }k=1, \dots, n \mbox{ (the superdiagonal)}
\\ &
v_{k-1,k} = \mu (n - k + 1) &&\mbox{ for }k=1, \dots, n  \mbox{ (the subdiagonal)}
\end{aligned}
\end{equation}
is required.
%
The determinants $\Delta_n$ with minimal indices $n=0,1,2,\dots$
arising  if $\llll=-1,$ $-2,$ $-3,\dots $
look as follows
\begin{equation}\label{Deltas}
\begin{aligned}
\Delta_0=\lambda,\;\Delta_1=\lambda^2-\lambda-\mu^2, 
\Delta_2= \lambda^3 -4\lambda^2-4\lambda\mu^2+4\lambda+4\mu^2,\dots 
\end{aligned}
\end{equation}
If some of them vanishes then the corresponding polynomial from the sequence
\begin{equation}\label{polyEs}
\begin{aligned}
&
E_{-1}=1,\;
E_{-2}= \mu z - \lambda +1,\;
\\&
E_{-3}= -2\mu^2z^2 + 2 (\lambda - 2)\mu\, z -\lambda^2 + 4\lambda + 2\mu^2 - 4,\dots
\end{aligned}
\end{equation}
solves \Eq\eqref{sDCHE},

If an integer $\llll > -1$ then
the latter equation has no polynomial solutions.
However the transformation
\begin{equation}\label{reversignofell}
E(z) \to \tilde E(z) = e^{\mu(z+1/z)}E(-1/z)
\end{equation}
converts any its solution to solution of the similar equation with the same  $\lambda$ and $\mu$
but with value of the parameter $\llll$ of the opposite sign. Accordingly,
as the result of transformations of solutions \eqref{polyEs},
 the functions
\begin{equation}  \label{quasipoly}
\begin{aligned}[t]
&
E_{0}=e^{\mu(z+1/z)}
,\;
 E_{1}
=
z^{-1} e^{\mu(z+1/z)}\big((1 - \lambda)z - \mu\big)
,\;
\\&
E_{2}
= [-] z^{-2}e^{\mu(z+1/z)}\big((\lambda^2 - 4\lambda - 2\mu^2 + 4) z^2  + 2 (\lambda - 2)\mu\,z  + 2\mu^2\big)
\dots
\end{aligned}\hspace{-2em}
\end{equation}
verify 
\Eqs\eqref{sDCHE} with $\llll=0,1,2,\dots$, respectively, provided the determinant
from the list
\eqref{Deltas} with the same subscript vanishes.
A core of the above and similar more complex formulas is some polynomial.
Hence they can be named quasipolynomial solutions.

The explicit solutions to \Eq\eqref{overdamped RSJ} constructed from the above  $E_0$ and $E_1$ to \Eq\eqref{sDCHE}
look as follows:
\begin{equation}\label{polycases}
\begin{aligned}[c]
\phi=\phi_0=&\; \half\pi + \omega\,\akatau\quad \mbox{for }A=1,\: B=\omega;
\\
\phi=\phi_1=&\; \Ratio{3}{2}\pi + 2\omega\,\akatau
+2\arctan{ \sin\omega\akatau  \over \sqrt{1+2\omega} -\cos\omega\akatau  }
\\ &\;
\mbox{for }A=\sqrt{1+2\omega},\: B=2\omega,
\\[-1ex] &\hphantom{.}\hspace{-4em}\mbox{\it etc}
\end{aligned}\hspace{-0em}
\end{equation}


As it has been noted
the family of equations \eqref{sDCHE} is equivalent to a proper subset of so called double confluent Heun equations
which is a subfamily of the family of Heun equations \cite{SW},\cite{SLS}.
One can find in the literature several representations of the latter but the most convenient here one is
given in \Src\cite{DIL} (its operator is displayed in the last formula in page 34). It reads
%
\begin{equation}
                   \label{DCHE genuine} %
0=
\ddrmG
+\big(\akat \,\akaz^{-2} + (2-\chi_0)\akaz^{-1} - 1 \big)
\drmG
+ \big( \half(\chi_\infty+\chi_0 - 1)\akaz^{-1} -c\,\akaz^{-2}\big)
\rmG
\end{equation}
Here $\akat,\chi_0,\chi_\infty,c$ denote arbitrary constants and $\akaz$ is the free complex variable
serving argument of the function $\rmG=\rmG(\akaz)$, the prime "\,$\drmchar{\mathstrut}$" denoting derivative with respect to $\akaz$.

It has to be taken into account that
in spite of the letter coincidence
the parameter $\akat$ from \Eq\eqref{DCHE genuine}
is not related in a meaningful sense to the dimensional variable $t$
from \Eq\eqref{RSJ or RCSJ}
formalizing the physical concept
of time.
These and some similar notations using common letters in application to distinct meanings
are in all cases traditional and it seems to be the best to follow
them
manifesting distinctions in meaning
by the distinction in imaging of the corresponding glyphs ensured by using of distinct available fonts.
In particular,
one could say that $t$ and $\akat$ are the same letter but distinct characters to be distinguished
by their meaning that has to be taken into account when interpreting formulas they are involved in.

Quite similarly, the symbol $z$ first appeared in \Eq\eqref{Riccati} and the symbol $\akaz$ from \Eq\eqref{DCHE genuine}
are to be considered distinct. They both serve free complex variables but differ although are related.
Such a relation will be specified soon.

The following linear homogeneous second order differential equation
%
%
%
%
%
\begin{equation}
                     \label{DCHE deformed}  %
\hphantom{.}\hspace{-2em}
\begin{aligned}[t]
0=
\; 
&\ddrmG+
\Big( \akat\,\akaz^{-2}+(2-\chiup_0)\akaz^{-1} -1 - {1\over \akaz-\bflam}\Big)
\drmG
\\[-1ex]
& \hspace{-2em}
+\akaz^{-2}\Big( \half (\chiup_0+\chiup_\infty - 1) (\akaz-\bflam)
 -\bflam^2\bfmu^2 +( \bflam^2 + \chiup_0\,\bflam - \akat)\,\bfmu +{\bflam^2\bfmu \over \akaz-\bflam}
\Big)
\rmG
\end{aligned}\hspace{-3em}
\end{equation}
is named the deformed double confluent Heun equation (its differential operator is displayed in the first formula in page 35 of \Src\cite{DIL}).
Here the unknown $\rmG$ is a function of the same complex variable $\akaz$ which has appeared in \eqref{DCHE genuine}, $\rmG=\rmG(\akaz)$.
The constant parameters $\akat,\chi_0,\chi_\infty$ refer to the eponymous ones from there while the parameters
$\bflam,\bfmu$ are new.
This is another case similar to the above one when
it should be emphasized that they has nothing in common with parameters
$\lambda$
and $\mu$
introduced earlier in \Eqs \eqref{AB vs lamu} and \eqref{sDCHE},\eqref{constconstr}, respectively.
The coincidence of notations borrowed from different sources is here occasional.
Hence the added distinction in characters used for their imaging.

Concerning the specifier `deformed' entering name of \Eq\eqref{DCHE deformed}, let us mention that in case $\bflam=0$
the equation takes the following form
\begin{equation}
                   \label{DCHE to deform} %
0=
\ddrmG
+\big(\akat\,\akaz^{-2} + (1-\chiup_0)\akaz^{-1} - 1 \big)
\drmG
+ \big( \half(\chiup_\infty+\chiup_0 - 1)\akaz^{-1} -\bfmu\,\akat\,\akaz^{-2}\big)
\rmG
\end{equation}
This equation arises from the equivalent equation \eqref{DCHE genuine} upon the following
replacements modifying its constant parameters
\begin{equation}
\chi_0 \leftleftharpoons \chiup_0+1, \;
\chi_\infty \leftleftharpoons \chiup_\infty - 1,\;
c \leftleftharpoons  \bfmu\,\akat
\end{equation}

Given a continuous curve in the space of constant parameters of \Eq\eqref{DCHE deformed} containing the point at which $\bflam=0$,
one may name deformation of the corresponding \Eq\eqref{DCHE to deform}
the set of equations arising along that curve. 
Among these deformations, special ones named monodromy preserving or isomonodromic are distinguished \cite{JMU}.
(The adaptation of the effect of isomonodromicity to the present context will be discussed below).
On the corresponding curve, the parameters $\chiup_0,\chiup_\infty$ do not vary
while the remaining $\akat,\bflam,\bfmu$ turn into  functions of a free variable.
To be specific, it is $\akat$ which plays the latter role 
while $\bflam $  and $ \bfmu $ become its functions of which it is claimed to obey the following system of the \textit{coupled\/} Riccati equations%
\footnote{although this is of course not a pair of genuine Riccati equations}
\begin{equation}     \label{lamuEqs}
\begin{aligned}
\dbflam =&\; 1 + \akat^{-1}\left(1 - \chiup_0 + (2\bfmu - 1)\,\bflam\right)\bflam ,
\\
\dbfmu = & \;
\akat^{-1}
\left(
-\half(\chiup_\infty +\chiup_0 - 1) + (\chiup_0  - 1 +2\bflam)\,\bfmu - 2\bflam\,\bfmu^2
\right)
\end{aligned}
\end{equation}
Here and in what follows `the accent'  ``\,$\dbf{\vphantom{\bfmu}}$''
denotes derivative with respect to $\akat$ and existence of $\akat_0>0$ such that $\bflam(\akat_0)=0$
is assumed.

\Eqs{\,}\eqref{lamuEqs} arise as Hamilton's equations for the Hamiltonian
%
\begin{equation}
                                 \label{Hamiltonian} %
\begin{aligned}[t]
{\mathcal H}(\bflam,\bfmu,\akat)&\;=
 \akat^{-1}\big(\bflam^2\bfmu^2-(\bflam^2 + (\chiup_0 - 1)\bflam - \akat)\,\bfmu
\\&\hspace{5.4em}
 + \half(\chiup_\infty+\chiup_0-1)\bflam\big)
\end{aligned}
\end{equation}

Upon elimination of $\bfmu$ from \Eqs\eqref{lamuEqs}, one obtains that
\begin{equation}
                              \label{eq. PIII'} %
    \ddbflam
    =
   {\dbflam^2 \over \bflam}
-{\dbflam \over \akat}
-{\chiup_{\infty}\,{\bflam}^2 \over \akat^2}
+{{\bflam}^3\over \akat^2}
+{\chiup_0 \over { \akat}}
-{1\over{\bflam}}
\end{equation}
This is the Third \Pain{} equation (the formula is equivalent to combination of the first and third lines of \Eq(5.6) from \Src\cite{DIL}).

It has to be noted that the system of Hamilton's equations
\begin{equation}   \label{eps = -1}
\begin{aligned}
\dbflam =&\; -1 + \akat^{-1}\left(\chiup_0 + 1 + (2\bfmu - 1)\,\bflam\right)\bflam,
\\
\dbfmu = & \;
-\akat^{-1}
\left(
          \half(\chiup_\infty - \chiup_0 - 1) + (\chiup_0 + 1 - 2\bflam)\,\bfmu + 2\bflam\,\bfmu^2
\right)
\end{aligned}
\end{equation}
inferable from the similar Hamiltonian
\begin{equation}
\begin{aligned}[t]
{\mathcal H}(\bflam,\bfmu,\akat)&\;=
 \akat^{-1}\big(\bflam^2\bfmu^2- (\bflam^2 - (\chiup_0 + 1)\bflam + \akat) )\,\bfmu
\\&\hspace{5.4em}
 + \half(\chiup_\infty - \chiup_0 - 1)\bflam\big)
\end{aligned}
\end{equation}
yields upon elimination of   $ \bfmu $ the same equation \eqref{eq. PIII'}.

It is known that the fulfillment of \Eq\eqref{eq. PIII'} is invariant with respect to the combined replacements
\begin{equation}
{\bflam}( \akat)\! \leftleftharpoons \akat \,\bflam(\akat)^{-1}
,\;
\chi_0 \rightleftharpoons \chi_\infty
\end{equation}
(these are equivalent to a particular case of the transformations $T_2$ given in \Src\cite{GLS}, see \Eqs(29.18) therein). 

Let us now reduce the
 pair 
of constants parameters $ \chiup_0, \chiup_\infty $
configuring \Eq\eqref{DCHE to deform}
 to a single one by means of the replacements
\begin{equation}\label{param reduction}
\chiup_0 \leftleftharpoons - \llll,\;\chiup_\infty \leftleftharpoons -\llll -1
\end{equation}
and replace simultaneously  the symbols   $\rmG$ and $\bfmu$
by the symbols
$\bfG$ and $\bfupmu$, respectively.
Then we obtain the equation
%
\begin{equation}
\label{J-Heun}
0=
\ddbfG
+\left(
{\akat\over \akaz^2}+ {\llll + 1 \over \akaz}   - 1
\right)\dbfG
-\left(
{\llll + 1  \over \akaz }
+{ \akat\,\bfupmu \over \akaz^2  }
\right)\bfG
\end{equation}
which is therefore a particular case of \Eq\eqref{DCHE to deform}.
Its  subsequent invertible transformation
by means of the rescaling of the  complex variable $\akaz$
in conjunction with natural redefinition of the unknown function  in accordance with the following template 
\begin{equation}\label{E2G and in reverse}
\bfG(\akaz) = E(z) \mbox{ for } z=\akat\,\akaz
\end{equation}
convert \Eq\eqref{J-Heun}  to \Eq{\,}\eqref{sDCHE} with $z$-independent parameters taking the following values 
\begin{equation}\label{parms generating}
\lambda =  -\akat\, \bfupmu,\; \mu =\sqrt{\akat}
\end{equation}
The following consequence of \Eq\eqref{constconstr} and \Eqs\eqref{parms generating}
\begin{equation}\label{om determination}
\omega =\half\left((1-\bfupmu)\akat   \right)^{-1/2}\!\!
\end{equation}
has to be accentuated.

On the other hand, \Eq\eqref{J-Heun}, as a particular case of \Eq{\,}\eqref{DCHE to deform},
possesses an own ``deformed counterpart'' contained in the family \eqref{DCHE deformed}.  It reads
\begin{equation}
\label{J-Heun deformed}
\hphantom{.}\hspace{-4em}
\begin{aligned}[t]
0=&\;
\ddbfG
+\left(
{\akat\over \akaz^2}+ {\llll + 2 \over \akaz} - 1 - {1 \over \akaz - \bfuplam}
\right)\dbfG
\\
&\hspace{1.4em}
-{1\over\akaz^2}
\left(
(\llll + 1)(\akaz - \bfuplam)
+
\left(\akat+ \big(\llll - \left(1 - \bfupmu\right)\bfuplam\big)\bfuplam\right)\!\bfupmu
- {\bfuplam^2\bfupmu \over \akaz - \bfuplam}
\right)\bfG
\end{aligned}\hspace{-3em}
\end{equation}
In case of zero $ \bfuplam $ this equation coincides with \Eq\eqref{J-Heun} becoming therefore equivalent to
\Eq{\,}\eqref{sDCHE} with parameters determined by the formulas \eqref{parms generating}.
\Eq{\,}\eqref{J-Heun} and its above `deformation' are close, provided $ |\bfuplam| $ is small
and $\akaz$ is kept outside some undiminishing neighborhood of zero.

 An isomonodromic deformation of \Eq{\,}\eqref{J-Heun} making use of \Eq{\,}\eqref{J-Heun deformed}
is realized in terms of the curve in the space of parameters of the latter
by means of regarding of $\bfuplam$ and $ \bfupmu $  as functions of the third parameter $\akat$.
They have to obey the corresponding adapted version of
\Eqs{\,}\eqref{lamuEqs}, i.e.\ the equations
\label{eqs for bfuplam and bfupmu}
\begin{eqnarray} \label{eq for bfuplam}
\dbfuplam &=& 1  + \akat^{-1}(\llll + 1 + (2\bfupmu - 1)\,\bfuplam)\,\bfuplam,
\\               \label{eq for bfupmu}
\dbfupmu &=&
\akat^{-1}
\left(
\llll + 1 +  2\bfuplam\,\bfupmu
\right)(1 - \bfupmu)
\end{eqnarray}
Upon elimination of $\bfupmu $
they yield the decoupled equation
%
\begin{equation}
                   \label{sPIII'} 
    {\ddbfuplam }
    =
   { \dbfuplam^{\!2} \over \bfuplam}
-{{\dbfuplam }\over \akat}
+{(\llll+1)\,{\bfuplam}^2 \over \akat^2}
+{{\bfuplam}^3\over \akat^2}
-{\llll \over { \akat}}
-{1\over{\bfuplam}}
\end{equation}
for the function $ \bfuplam(\akat) $.
This is a particular case of 
\PIIIeq{}
\eqref{eq. PIII'} 
corresponding to the reduction \eqref{param reduction} of its parameters.
Similar elimination of $\dbfuplam$ yields the 
decoupled
equation
%
\begin{equation}\label{sPIII' for bfupmu}
\ddbfupmu =
    { \dbfupmu^2 (1 - 2\bfupmu) \over 2(1 - \bfupmu) \bfupmu }
   -  {\dbfupmu  \over \akat}
   +  {2  (1 - \bfupmu)\bfupmu  \over \akat}
   - { (\llll + 1)^2 \over2 \akat^2 }{(1 - \bfupmu) \over \bfupmu }
\end{equation}
for the function $ \bfupmu(\akat) $.

In addition to functions $\bfuplam(\akat), \bfupmu(\akat)$ there exist their counterparts which we denote $\dualam(\akat), \duamu(\akat)$
(to be named, e.g., `mirrored [boldface] $\lambda$, $\mu$') with the following rules of the bidirectional transformations:
\begin{eqnarray}
\dualam \!\!\!&=&\!\!\! \akat\, \bfuplam^{-1},\; \duamu = \akat^{-1}\bfuplam^2(1-\bfupmu)
                              \label{forward}
\\[-2.5ex]
 \mbox{\raisebox{2.0ex}{\hspace{-2.0ex}\llap{$\rightharpoondown\hspace{0.55ex}$} }}
\llap{$\bfuplam\,\dualam  = \akat \hspace{0.6em}\Big\langle$ \hspace{1.1em}}\nonumber
\\[-2.5ex]
 \mbox{\raisebox{-0.2ex}{\hspace{-2.0ex}\llap{$\rightharpoonup\hspace{0.5ex}$} }}
\bfuplam \!\!\!&=&\!\! \! \akat\, \dualam^{-1},\; \bfupmu =  1 - \akat^{-1}\dualam^2\duamu
                              \label{backward}
\end{eqnarray}
The new parameters obey the equations (following from \Eqs{\,}\eqref{eq for bfuplam} and the above transformation rules)
\begin{equation}\label{eqs for dualammu}
\begin{aligned}
\ddualam =&\:
-1 - \akat^{-1}\big(\llll  -  (2\,\duamu - 1)\dualam\big)\dualam,
\\
\dduamu = &\;
 \akat^{-1}
\big(
\llll +   2\dualam\,(1 - \duamu)
\big)\duamu
\end{aligned}
\end{equation}
These equations are a special instance of \Eqs\eqref{eps = -1}.
The decoupled equations for $ \dualam(\akat) $  and $ \duamu(\akat) $ implied by \Eqs{\,}\eqref{eqs for dualammu}   read
\begin{eqnarray}
                 \label{sduaPIII'}
    \dddualam
    &\!\!=\!\!&
   { \ddualam^{\!2} \over \dualam}
-{\ddualam \over \akat}
+{\llll\,{\dualam}^2 \over \akat^2}
+{\dualam^3\over \akat^2}
-{ \llll+1 \over { \akat}}
-{1\over \dualam},
\\ 
  \label{eq for duamu}
\ddduamu
    &\!\!=\!\!&
    { \dduamu^2 (1 - 2\duamu) \over 2(1 - \duamu) \duamu }
   -  {\dduamu  \over \akat}
   -  {2  (1 - \duamu)\duamu  \over \akat}
   + { \llll^2 \over2 \akat^2 }{\duamu \over 1 - \duamu}
\end{eqnarray}
\Eq{\,}\eqref{sduaPIII'} represents another special instance of Third' \Pain{} equation \eqref{eq. PIII'}
while \Eq{\,}\eqref{eq for duamu} has much in common with \Eq{\,}\eqref{sPIII' for bfupmu}.

\subsection{Phase-lock in evolution of state of a Josephson junction}

\hphantom{ }\\[1ex]
While
\Eq\eqref{momentaryV} determines
the momentary voltage $V(t)$ across Josephson junction via the rate of variation of the phase difference $\varphi(t)$,  
the  averaged voltage
$\langle V\rangle_t = t^{-1} \int_0^{t}  V(\tilde t)\,d \tilde t$ 
in case of 
the time lapse $t$ much greater than the period 
$T $ 
of the AC-constituent of the bias current 
is usually represented by its limit with $t=N T$ and integer $N\nearrow\infty.$ \Eq\eqref{momentaryV} then implies that
$\langle V\rangle_\infty= \lim_{N\to\infty}(\hbar/(2e))(T\,N)^{-1}\varphi(T\,N)$. 
In terms of the normalized time variable $\akatau$ expressed via $t$ by the first of the formulas \eqref{defs of tau et al},
the phase $\varphi(t)$ replaced with $\phi(\akatau)$, and the period $\Tau$ of variation of $\akatau$ equivalent to the period $T$
of variation of $t$ determined by periodicity of the bias $I(t)$, 
the above reduction takes the form $ \langle V\rangle_\infty=\lim_{N\to\infty}I_c/(G\,\Tau N)\phi(\Tau\,N)$
which is equivalent to
\begin{equation}        \label{V observed}
\langle V\rangle=
\langle V\rangle_\infty={\omega I_c\over G} \,\rotnum, \mbox { where }\omega={2\pi\over\Tau} \mbox{ and }\rotnum = \lim_{N\to\infty}{\phi(N \Tau)\over 2\pi N}.
\end{equation}
The quantity $\rotnum $ is a function of model constant parameters
which proves to be 
independent of the particular choice of the solution $\phi$ to \Eq\eqref{overdamped RSJ} 
used for its determination in terms of the above limit.
It is named \textit{ the rotation number}.
The factor of $(2\pi)^{-1}$ in the last definition playing role of a kind of normalization 
is related to invariance of the equations \eqref{RSJ or RCSJ} and
\eqref{overdamped RSJ} with respect of the shift of unknowns by $2\pi$ allowing to consider them angular by nature
and then to evaluate against the $2\pi$-scale in units of `the number of revolutions' (i.e.\ the rotation number) 
with particular attention to its integer values.
The latter
case is most important since in presence of the phase-lock the rotation number is always integer \cite{BKV}, \cite{IRF}.

The problem of efficient description of the dependence of the rotation number
on the parameters characterizing
Josephson junction and external impact on it
is of notable importance.
In particular,
in accordance with relation \eqref{V observed}, its solution would enable one to predict the value of voltage across the junction.
Moreover, the theory and observations show that the rotation number is always well defined and, under certain conditions, proves to be highly stable
against parameter perturbations. 

\begin{figure}[h]
\scalebox{1.0}{
\includegraphics{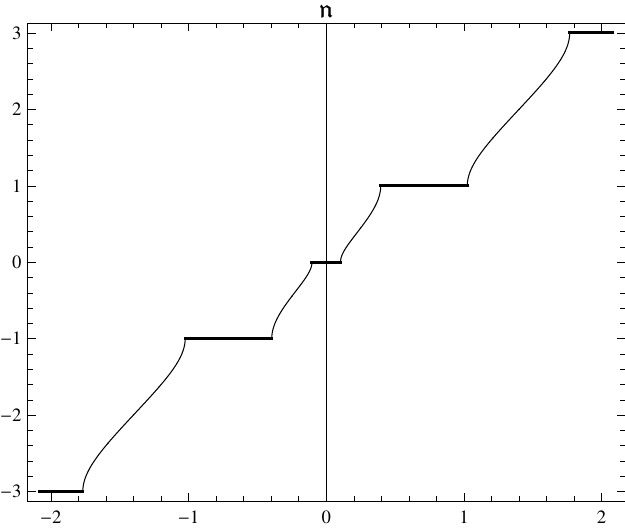}
}
\caption{
\small
The graph representing
a typical 
dependence of the rotation number $\rotnum$ proportional to the avaraged voltage $\langle V\rangle$
for the model utilizing \Eq\eqref{overdamped RSJ}
on the bias parameter $B$ proportional to DC through junction
(IV-curve) in case of their small and moderant values. Other parameters are fixed as follows: $A=1.3,$ $\omega=0.7$.
}\label{fig1}
\end{figure}

In framework of the model based on the equation \eqref{overdamped RSJ} the parameters to be taken into account are the constants
$A,B,\omega$. Some characteristic features of the dependence of the rotation number on them can be learned
with the help of a body of
graphs plotting  value of $ \rotnum $ fixing  two among the above parameters and varying the remaining one.
These often are the normalized AC amplitude $A$ and the normalized frequency $\omega$  which are fixed while the role of free variable
is played by the normalized DC bias $B$.
 A typical graph arising in this way is shown in \Fig\ref{fig1}. 
\noindent
In principle, it could be extended well to the left and to the right without limitation, its stair-like form
being there observed as well.

Specifically, the curve of the graph in \Fig\ref{fig1} contains a sequence of horizontal straight segments of varying lengths but uniform vertical distribution
occupying sequentially
all the integer levels. The opposite boundary points of nearby straight segments are connected by
monotonously tilted curves of left-to-right positive and negative  curvatures which are
smooth everywhere except the boundary points where they become vertical.
The central symmetry of the whole picture with respect to the zero center of coordinates is observed.

Up to appropriate assigning of units
the graph in \Fig\ref{fig1} represents the I-V (direct current -- averaged voltage) dependence
which a circuit with overdamped Josephson junctions exhibit.
The horizontal segments it contains are named  Shapiro steps (see \cite{Sh}, \cite{SJH}, \cite[section 10.6.2]{MK}). The voltage across junction
remains strictly constant as long as the external impact is keeping its state within a Shapiro step
(and the frequency $\omega$ hasn't changed).

Horizontality of Shapiro steps and their integer levels mean that asymptotically the incrementation of any solution to \Eq\eqref{overdamped RSJ}
across the lapse of normalized time $\tau$ equal to  $\Tau=2\pi\omega^{-1}$
amounts to a fixed integer multiple of $2\pi$.
The criterion of manifestation of such a property 
can be reduced to condition of fulfillment of the inequality
\begin{eqnarray}
|\Lambda| &>& 1,
\end{eqnarray}
where $\Lambda$ (which is always real) has been defined in \eqref{mprediscr}.
If this condition is met the rotation number can be found as the result of division  by $2\pi$ of
the quantities equated in \eqref{increment}.
In the present case the asymptotic approximating \eqref{V observed}
to the rotation number is realized on a finite interval of variation of $\tau$
in view of existence of  solutions  periodic \modtwopi.

On the other hand,
it holds $|\Lambda|<1$ if a point of the graph is chosen between Shapiro steps.
Then the rotation number is not integer.
Its fractional part
is defined (for $B>0$)
by the formula
\begin{equation}\label{inbetween}
\rotnum   \stackrel{\mbox{\tiny\sf mod}\:1}{ =\; }
{1\over 2} - {(-1)^{\rotnum _p}\over \pi}\arctan{\Lambda\over \sqrt{ 1-\Lambda^2}}
\end{equation}
The corresponding integer part is equal to the rotation number $\rotnum_P$ associated with the preceding Shapiro step.
Starting with the zeroth one observed around the center $B=0$ and moving to the right,
the quantities $\rotnum_P$ combine in a general case to the sequence $0,1,2,3\dots$ The resulting graph appears to be continuous (connected)
but not smooth at boundary points of Shapiro steps.
The part of figure with negative $B$ is provided by application of the point symmetry with respect the zero center.

\begin{figure}[h]
\scalebox{1.0}{
\includegraphics{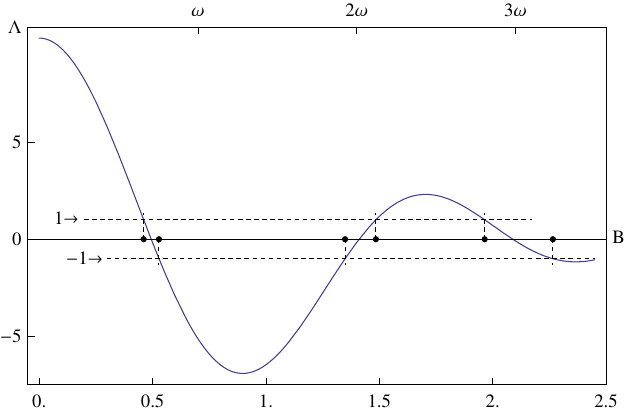}
}
\caption{ 
\small
The dependence of value of the functional $\Lambda$ defined by the formula \eqref{mprediscr}
on the parameter $B$
in case of
$A=1.3$ and $\omega=0.7$.
The bold points on the horizontal coordinate axis indicate the values of $B$ yielding $|\Lambda|=1$.
}\label{fig2}
\end{figure}

Specific details of the picture
imaged in \Fig\ref{fig1} can be deduced from certain dependencies associated
with the quantity $\Lambda$ defined in \eqref{mprediscr}.
In the case of fixed  $A=1.3$ and $\omega=0.7$ we have to consider
the graph representing $\Lambda$ as a function of $B$. %
It is shown below in \Fig\ref{fig2}.
Its intersections
with levels $\pm1$ are here indicated by projections to abscissa.
The corresponding values of $B$
represent the horizontal coordinates%
\footnote{These are
{\tiny
$|\!\!\rightarrow
{0.46060\!\dots}
\leftrightarrow {0.5280\!\dots}
\longleftrightarrow{1.3490\!\dots}
\leftrightarrow{1.4850\!\dots}
\longleftrightarrow{1.96488\!\dots}
\leftrightarrow{2.2659\dots} \leftarrow\!\!|$
}
\label{ftn1}
}
 of the boundary points for Shapiro steps
on the graph shown in \Fig\ref{fig1}.

The substantial distinction of behavior of the rotation number on the parameters (this is $B$ in the case we have outlined)
on different parts of the graph in \Fig\eqref{fig1}
is closely related to  asymptotic properties of subsets of solutions to \Eq\eqref{overdamped RSJ}.
It is known that the latter equation, as well as some its generalizations,
admits two distinct perturbatively stable modes of asymptotic behavior and a number of intermediate ones.

In the first case
the set of solutions starting from the segment $[0,2\pi)$, i.e.\ defined by initial values
obeying inequalities
$0\le\phi(0)<2\pi$,
contains exactly two ones which are periodic with the period $\Tau=2\pi/\omega$ up to a common constant increment equal to certain integer multiple of $2\pi$.
One may say that they are periodic \modtwopi{} or \modtwopi-periodic.
Among them, there is a solution which is stable with respect to perturbations of the initial value $\phi(0)$
while another one is not.

All the other nearby solutions of the same equation exponentially converge to the stable \modtwopi-periodic solution which serves them an attractor.
The unstable \modtwopi-periodic one plays role of a repeller.
As a consequence, asymptotically, all solutions gain the property of  \modtwopi-periodicity
up to ever-decreasing discrepancies.
Such a common specialty of evolution of solutions to \Eq\eqref{overdamped RSJ} is named phase-lock.
The subset of the space of its constant parameters in which phase-lock is observed is named phase-lock area. It is open.

A demonstrative example depicting
the flow of integral curves for \Eq\eqref{overdamped RSJ} in case of phase-lock is given in \Fig\ref{fig14} (Appendix, page \pageref{fig14}).

The second perturbatively stable mode of collective asymptotic behavior of solutions to \Eq\eqref{overdamped RSJ}, regarded as the property
of the very equation, is manifested, in contrast, in apparent absence of any stable periodicity and convergences.
There still exists a countable
subset of solutions starting, as above, from the segment  $[0,2\pi)$
such that $ \phi(\akatau + j\, k\, \Tau) \stackrel{\mbox{\tiny \modtwopi}}{ =\; } \phi(\akatau)  $
for some $\akatau\ge 0$, some integer $k$, and all integer $j=1,2,3\dots$.
However, there are arbitrarily weak perturbations which eliminate appearance of the noted property.
This also means that
for each solution from the complementary subset of solutions of a full measure
no sequences $\phi(\akatau + j\, \Tau)$, $j=0,1,2,3\dots$ contain pairs of members coinciding \modtwopi, i.e.\ all they manifest
no repetitions \modtwopi{} at all.
Besides, any such sequence generated by a single solution,
when get \modtwopi, proves to be chaotically distributed and dense in the segment $[0,2\pi)$.
Such discrete evolutions
can be named pseudo-chaotic since  apparent chaos in subsequent values of $\phi(\akatau)$ get at moments of `time' separated by the lapse $\Tau$
arises here
as a consequence of interplay of two continuous periodic ``processes'' whose frequencies are inconsonant. %

A telling example of result of computation illustrating
apparent disorder 
observed in evolution 
described by
solution to \Eq\eqref{overdamped RSJ} in case of the stable absence of the phase-lock in given in \Fig\ref{fig15} (Appendix, page \pageref{fig15}).

A continuous varying of constant parameters can induce a transition between two open subsets of the space of constant
parameters taking down the property of phase-lock or, conversely, finishing with its appearance. In such a process an intermediate state
manifested on the common boundary of the two above subsets has to meet. In particular, there are \modtwopi-periodic solutions
of two kinds (stable and unstable) in case of phase-lock while none such solutions exist
for parameters from the open part of the complementary subset of their space.
In the process of varying parameters in a way leading to disappearance of phase-lock,
\modtwopi-periodic solutions can not merely disappear.
They may however
come pairwise together
that leads to their merging taking place when the parameters varied reach the common boundary of distinguished subsets of the parameter space.
On it, there retains only one \modtwopi-periodic solution inheriting one side of the graph which is `stable' (attractive) while another one
remains `unstable' (repelling). Other two sides, existed before, attracting and repelling, have come together and annihilated.
The boundary points of Shapiro steps shown in \Fig\ref{fig1} correspond to instances of \Eq\eqref{overdamped RSJ} possessing such a peculiarity.

A somewhat specific example of values of parameters also providing such a degenerate form of the phase-lock is specified in comments
to  \Fig\ref{fig16} serving also its useful illustration (Appendix, page \pageref{fig16}).
This case is also worth noting because of incorporation of one of the explicit solutions to \Eq\eqref{overdamped RSJ}
generated from explicit solutions to \Eq\eqref{sDCHE} --- a quasipolynomial solution.
The collections of parameters corresponding to all instances of
\Eq\eqref{overdamped RSJ}
obtained in this way belong to boundaries of phase-lock areas. At the same time, the latter in not exhausted by the former
since, in particular, for all quasipolynomial solutions it holds $B=\omega\times\!$ an integer that is not here required.

\begin{figure}[h]
\scalebox{1.0}{
\includegraphics{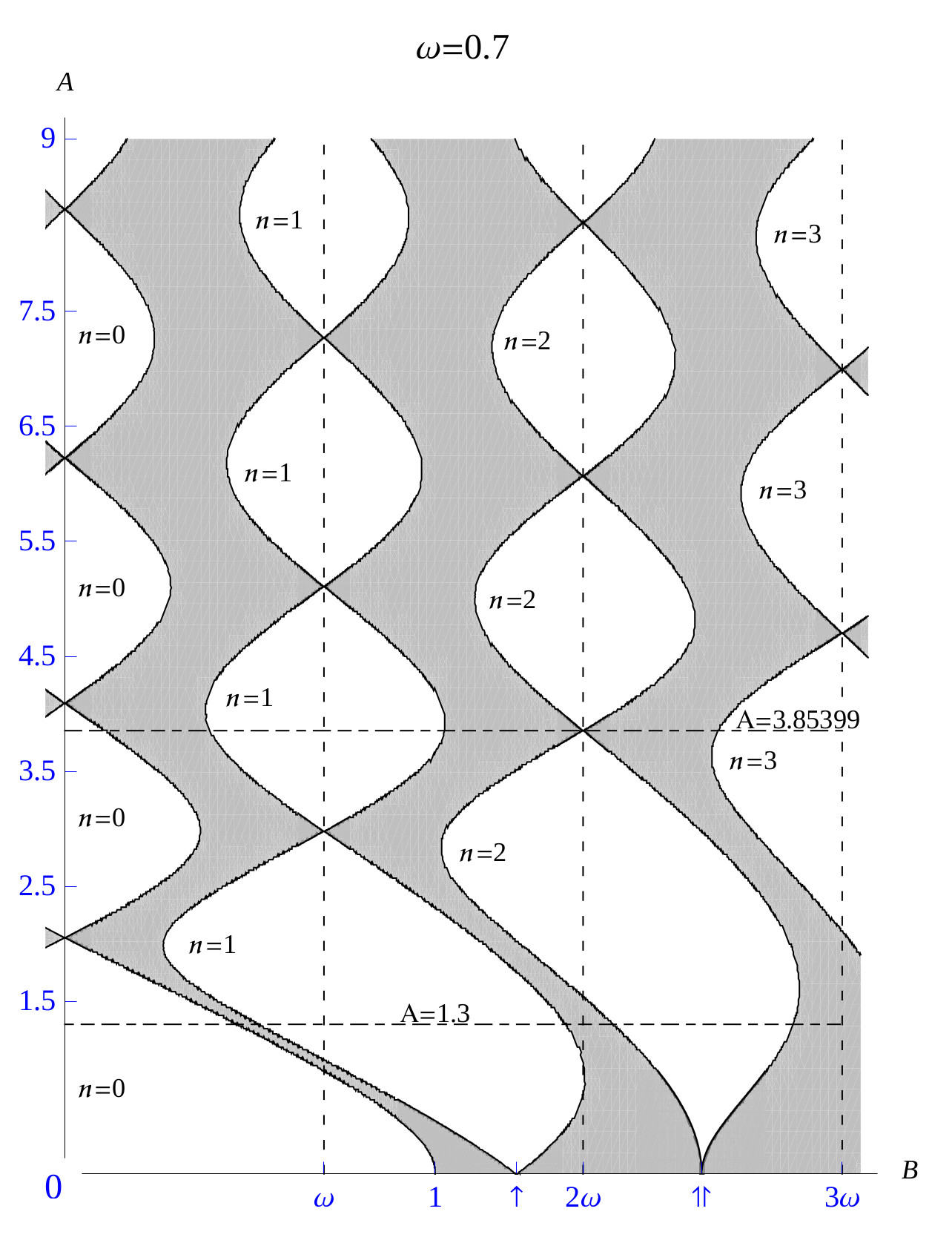}
}
\caption{
\small
A part of the section of the space of constant parameters of \Eq\eqref{overdamped RSJ} by the plane $\omega=0.7$.
White ("colorless") points beget equations from the noted family which possess the property of phase-lock
while pseudo-chaotic evolution characterizes
equations associated with points colored in gray.
The marks 
$\uparrow$ and $\upharpoonleft\!\!\upharpoonright$
indicate the shifts $\sqrt{1+\omega^2}$ and  $\sqrt{1+(2\omega)^2}$, respectively,
of the first two growth points.
When unboundedly extended, the geometric matter outlined is mirror symmetric with respect to both coordinate axes.
The rotation numbers $\rotnum$ characterizing phase-lock subareas are indicated.
The two broken horizontal lines sketch out the sections $A=\;$const for levels
$1.3$ (generic) and $3.85399\!\dots$ (exceptional)
of which the latter passes through the constriction of the phase-lock subarea of the rotation number 2.
}\label{fig3}
\end{figure}

The graph shown in Fig. \ref{fig1} and analogous ones which would arise in case of other values of the constant parameters give some information
on the situating of phase-lock areas in the parameter space
but are inefficient for reaching a satisfactory insight.
The geometry of phase-lock areas is more fully demonstrated
by 2-dimensional graphs in which only one constant parameter (usually $\omega$) is fixated
while  other ones (we will use $A$ and $B$) are freely varied in a domain as wide as possible.
One of the such pictures can be seen in \Fig\ref{fig3} (page~\pageref{fig3}). 

In it, a part of the plane of values of the parameters $B$ and $A$  around the coordinate center
falling into the quadrant of their positive values is deployed.
Since the full image involving all the four quadrants is mirror symmetric
with respect to the both coordinate axes (as well as point symmetric at zero)
it complete form can easily be reconstructed from the part presented.
(It  should be borne in mind that
the numerical characteristics attributed to figures situated in different quadrants
and related by such a symmetry
can prove to be not identical but differing by opposite signs.)

The image consists, basically, of two colors. Namely, some points are gray and some are white (`colorless').
Additionally, for easy visibility, the black color is used for indication of boundaries.

The pairs of values of parameters $B,A$ retrieved from colorless places provide (for $\omega=0.7$ that is assumed throughout most of our numerical examples)
the specimens of \Eq\eqref{overdamped RSJ} describing evolution revealing phase-lock 
with external periodic impact encoded in the equation right hand side.
Accordingly,
asymptotic convergence \modtwopi{}
of their generic solutions
to certain \modtwopi-periodic ones takes place.
This means that the white pieces of the picture represent the subsets of the phase-lock areas. All they are open
and each of them is characterized by an integer rotation number introduced by the last formula of \Eqs\eqref{V observed}.
As to numerical evaluation, the rotation number at points of a phase-lock area can be determined as the result of
division of the right hand side of the formula \eqref{increment} by $2\pi$.

In case of
$B,A$ picked from the open part of the subset of the space of constant parameters complementary to the phase-lock area
which is colored in gray
the phase evolution does not manifest
on a long scale
any convergence or
self-reproduciblility which would be preserved under the impact of sufficiently small perturbations.
This case has been mentioned as pseudo-chaotic evolution.
In spite of such an apparent asymptotic unpredictability
these points are also endowed with unique rotation numbers which are manifested as a continuous function of the parameters
with non-integer values.
The latter can be determined by means of \Eq\eqref{inbetween}.
Some instances of the form of dependence of the rotation number on the bias $B$ in case of pseudo-chaotic evolution
are exhibited in \Fig\ref{fig1} by curved segments of the graph shown.

The relation of data borne by \Fig\ref{fig1} and \Fig\ref{fig3} can be displayed by means of a kind of overlapping of the former over the latter.
Computations yielding the graph in \Fig\ref{fig1} are carried out for $A=1.3$. The corresponding section along the level $A=1.3$ is depicted
as a horizontal broken line in \Fig\ref{fig3}. The vertical projections to the $B$-axis of its intersections with boundaries of the phase-lock areas
yield the values of abscissa for the boundaries of Shapiro steps shown in \Fig\ref{fig1}. See also footnote \ref{ftn1} (p.~\pageref{ftn1}).
The color of background points on the level $A=1.3$ in \Fig\ref{fig3}
enables one to determine 
whether the corresponding value of $B$ 
determines the point on a graph
from \Fig\ref{fig1} belonging to a flat Shapiro step
or it is situated on a curved segment inbetween two of them.
Thus while \Fig\ref{fig1} provides the quantitative data for a single value of $A$
the data provided by  \Fig\ref{fig3} is partially qualitative but at the same time it is augmented by possibility of consideration
of different values of  $A$ `at once'. In particular, the dependence of location and widths of Shapiro step on the parameter $A$
(i.e.\ on the intensity of AC power irradiating circuit with Josephson junction) can be predicted. 
It is seen, in particular, that Shapiro steps of
a certain level (associated with some given value of the rotations number)
can sometimes shrink
to zero width --- i.e.\ disappear.
Or conversely, one could search for conditions under which step width becomes maximal that is the case of interest from the viewpoint of applications.

It has also to be reminded that both representations \Fig\ref{fig1} and \Fig\ref{fig3} display geometry of
phase-lock area for a single fixed value of the frequency $\omega=0.7$.
For its moderate values similar images  appear to be not too different but in  the limiting cases of very large and, especially, very small $\omega$
(in fact, it suffices to get $\omega<0.2$) additional intriguing geometric relations become apparent  \cite{BG}, \cite{Gl2}.

\Fig\ref{fig3} illustrates the following features of phase-lock areas
for \Eq\eqref{overdamped RSJ} in the space $\mathbb{R}\times\mathbb{R}\times\mathbb{R}_+\ni(B,A,\omega)$ of its constant parameters.

In sections by the planes $\omega=\mathrm{const}>0$ the phase-lock area is divided into a sequence of finitely separated uniformly distributed subareas
extending in vertical direction (direction of variation of $A$)
which can be identified by subsequent integer values of the rotation number $\rotnum $ which are positive in the half-plane $B > 0$ and negative for $B < 0$.
The exception is the phase-lock subarea with zero rotation number, where $B$ can be negative, zero, and positive,
which is
skewered on the vertical coordinate axis and is mirror-symmetric with respect to it.
Going beyond the picture edges to the right and to the left, every  integer rotation number can be met.

In turn, a phase-lock subarea associated with given integer rotation number $\rotnum$ has the look of a (projection of a) kind of garland
stretched in vertical direction
and shifted in horizontal direction from the vertical symmetry axis by  $\rotnum \,\omega$.
Every such garland (phase-lock subarea)
consists of
the two-sided sequence of
slightly deformed lemon-like figures%
\footnote{see
\url{https://en.wikipedia.org/wiki/Lemon_(geometry)}
}
adjacent at corner points of their boundaries
whose combination
appears as the union of two smooth oscillating curves.
All the corner points of pairwise lemon concatenations
belong to the line $B= \rotnum \,\omega$ \cite{GKFS}
except the one falling on the horizontal coordinate axis.
Phase-lock subarea of zero rotation number again represents the exception: its intersection with abscissa is not a point but a segment.
The noted  points are named constrictions (of the corresponding  phase-lock subareas).
They embody conditions of degenerating (disappearance) of certain Shapiro steps.

The  lemons with one of the corner boundary points belonging to zero level line $A=0$ 
are more deformed as compared to others due to the pushing this point away off the center by the shift
$(\rotnum \,\omega)\sqrt{1+\rotnum ^{-2}\omega^{-2}}$ greater than the shift $\rotnum \,\omega$ characteristic for location of constrictions.
The noted corner points situated on the horizontal coordinate axis are named the growth points.

It is useful to introduce some identification mark for the above `lemons' constituting separate phase-lock subareas.
This is an integer
number named level which is equal to the serial number of given lemon above (then it is positive)
or below (respectively, negative) the abscissa axis. The levels of lemons steming from growth points are equal to $+1$ or $-1$.
Going upwards in the upper half-plane, they are being incremented by the unit in a natural way.
Accordingly, the shifts in the lower half-plane (not shown in \Fig\ref{fig3})  in opposite direction along the sequence of lemons of a given phase-lock subarea
are accompanied by the one by one level decrementations.

A defect of this way of lemon indication by integer rotation number and level is that there is no natural level for the lemon around
the coordinate center. The assignment to it zero value proves to be out the picture. It seems sometimes more correct to consider its level to be equal
to $+1$ but in other cases the level $-1$ seems to be more adequate. In particular, the levels of lemons adjacent to the one in question
from above  and from below (the latter is not shown in \Fig\ref{fig3}) are equal to $+2$ and $-2$, respectively.

We will also use sometimes a similar notion of levels for constrictions. Such a number is equal to the level of the lemon situated either underneath
(for the upper half plane) or above (for the lower one) for which the constriction is a corner point (and hence a point of the boundary).

As it has been noted,
there is one exception in the structure of constituents of the phase-lock area outlined above.
In the case of
zero rotation number
no growth point for the garland representing the corresponding subarea
exists. %
Instead, the whole segment \{$B\in(-1,1), A=0$\} belongs to it.
There is a lemon around it which is itself mirror-symmetric with respect to the coordinate axes
and one might say that it plays role of two lemons from phase-lock subareas with $ \rotnum  \not=0$ concatenated at the growth point
from above and from below,
the only point of the boundaries being situated on the coordinate axis in that case.
In case of zero rotation number the latter `blows up'
giving rise to the noted finite segment.

The constrictions serving the upmost and the lowest points of the closure of the lemon around the symmetry center are endowed with levels
$+1$ and $-1$, respectively

\begin{figure}[h]
\scalebox{0.8}{
\includegraphics{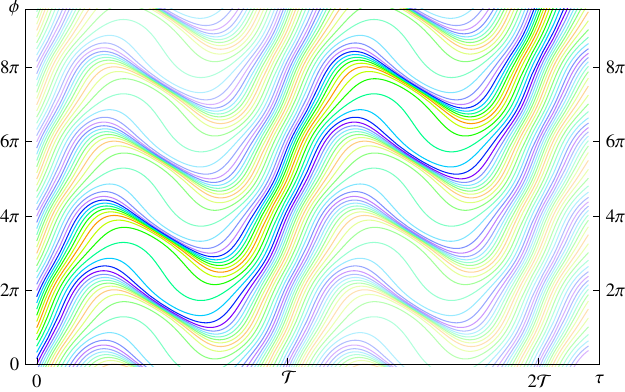}
}
\caption{ 
\small
A fragment of the flow
formed by integral curves for
\Eq\eqref{sDCHE} with parameters $\omega=0.7$, $B=2\omega$, $A=3.85399\dots$
corresponding to constriction on the phase-lock area shown in \Fig\ref{fig3}.
Each $2\pi$-long segment on the vertical coordinate axis $\akatau=0$ (not shown) 
contains 12 uniformly distributed starting points of graphs of solutions
$\phi(\akatau)$.
Upon unlimited continuation,
the invariance with respect to vertical shifts by $2\pi$
and horizontal shifts by $\Tau=2\pi/\omega$
takes place.
The rotation number demonstrated by the flow is equal to 2, i.e.,
in view of the noted symmetries, it holds
$\phi(\akatau+\Tau)= \phi(\akatau) + 2\cdot2\pi$ for all $\akatau$ and each solution $\phi$.
}
\label{fig4}
\end{figure}

A distinguishing property of parameters get from constrictions is as follows: all solutions to the corresponding \Eq\eqref{overdamped RSJ}
are
$\mbox{\small\sf mod}\: 2 \pi$-periodic with common increment across an interval of variation of their argument $\akatau$ of the length $\Tau$
equal to
$2\pi\,\rotnum $.
An example of the flow produced by this equation possessing the noted property is depicted in \Fig\ref{fig4}. 
The lowermost constriction of the level 1
situated
on the vertical shifted by $1.4=2\omega$
shown in \Fig\ref{fig3}
is there used.
Here the factor of 2 on the right determines the rotation number equal to the same integer, $\rotnum =2$.

Given fixed $\omega>0$, the values of parameters $A,B$ corresponding to constrictions of phase-lock area  
are 1-to-1 related to the points in the space of parameters $\lambda,\mu$ (the corresponding transformation is described by \Eq\eqref{AB vs lamu})
at which \Eq\eqref{sDCHE} admits a solution holomorphic at zero.
More generally, it has been shown in \Src\cite{BG} (Proposition 4.1) that the parameters of \Eq\eqref{overdamped RSJ} correspond to
a constriction if and only if the monodromy of the associated equation \eqref{sDCHE} is trivial.
This can only take place  if $\ell=B/\omega$ is non-negative integer
(the case of negative $\ell$ is handled by means of application of transformation \eqref{reversignofell} reversing sign of this parameter).

Moreover, if $E(z)$ is a solution to \Eq\eqref{sDCHE} holomorphic at zero (i.e.\ an entire function) then
the function
$\sharp E(z)$
obtained by means of the transformation
defined in \Eq\eqref{diez} and also verifying
the same equation, although singular at zero, is single-valued holomorphic in $\mathbb{C}\backslash 0$
and linear independent with $E(z)$. Accordingly, all solutions to \Eq\eqref{sDCHE}
which are linear combinations of $E(z)$ and $\sharp E(z)$ represent single-valued functions defined everywhere except (perhaps) at zero.
This means that the monodromy group
characterizing the space of solutions to
the equation considered is trivial. For generic (`for almost all') instances of \Eq\eqref{sDCHE}
the latter triviality is not observed,
i.e.\ the associated monodromy groups are non-trivial implying that either some or, more typically, all their non-trivial solutions cannot be single-valued
when considered
on the domain $\mathbb{C}\backslash0$.

\begin{figure}[h]
\scalebox{1.1}{
\includegraphics{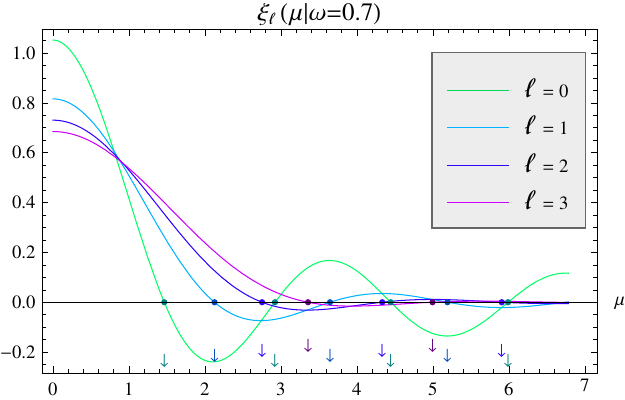}
}
\caption{ 
\small
Graphs of functions $\xi_\ell  $ defined by \Eqs\eqref{def of xi}, \eqref{withnotherarg} for $\omega=0.7$ and $\ell=0,1,2,3$.
Locations of their roots are indicated.
}\label{fig5}
\end{figure}

A solution to \Eq\eqref{sDCHE} holomorphic at zero (an entire function), if exists, is unique up to multiplication by a constant.
The necessary condition of its existence
is a non-negative integer value of $\ell$.
The sufficient condition is the vanishing of the function $\xi_\llll$ of the equation parameters
defined by \Eq\eqref{def of xi} which can also be represented in the form \eqref{withnotherarg}.
Besides, dealing with entire solutions, one may consider the sequence of separate $\xi_\ell,\;\ell=0,1,2,\dots$
The graphs of its first representatives are shown in \Fig\ref{fig5} (page \pageref{fig5}).
They constate the presence of desirable roots. Multiplying these values of $\mu$ by $2\omega$,
we obtain in accordance with \Eq\eqref{AB vs lamu} the set of values of the parameter $A$ used in \Eq\eqref{overdamped RSJ}
(other parameters are equal to $\omega=0.7$ and $B=\rotnum\,\omega=\ell\omega$).
They are given in table~\ref{tab1}.  
\begin{table}[t]\small
\begin{tabular}{rllll}
\ \\
\mbox{\hphantom{$\ell =0:$}\vphantom{ $\mbox{$A$} \atop \rotatebox{90}{=}$ }}
 & \multicolumn{1}{l}{$\mbox{$A$} \atop \rotatebox{90}{=}$}
 & \multicolumn{1}{l}{$\mbox{$A$} \atop \rotatebox{90}{=}$}
 & \multicolumn{1}{l}{$\mbox{$A$} \atop \rotatebox{90}{=}$}
 & \multicolumn{1}{l}{$\mbox{$A$} \atop \rotatebox{90}{=}$}
\\
$\ell =0:$ & 2.05265\!\dots &4.09182\dots&6.22422\dots&8.38662\dots
\\
$\ell=1:$ & 2.97808\dots&5.10662\dots&7.27032\dots &$\langle\,\cdots\rangle$
\\
$\ell =2:$ & 3.85399\dots&6.06800\dots&8.27050\dots&$\langle\,\cdots\rangle$
\\
$\ell =3:$  &4.70168\dots&6.99480\dots&$\langle\,\cdots\rangle$&$\langle\,\cdots\rangle$
\end{tabular}
\caption{\small
Values of the parameter $A$ inferred from roots $\mu$ of functions $\xi_\ell(\mu|0.7)$ (see \Fig\ref{fig5})
in accordance with the first of the formulas \eqref{AB vs lamu}.
}
\label{tab1}
\end{table}

In it, one finds the coordinates of all the constrictions of the phase-lock subareas within their parts fitted on \Fig\ref{fig3}.

The existence of entire solution to \Eq\eqref{sDCHE} is thus the exceptional property which is of notable interest.
Determination of conditions implying its possession can be can be carried out by means of investigation of $\xi$-function \eqref{def of xi}.

There is however yet another
way to do that which is worth a separate investigation.
It applies isomonodromic deformation \cite{JMU,JM1,JM2}, \cite{Sl}, \cite{Gl1}, \cite{T4} of the known instance of \Eq\eqref{sDCHE} admitting entire solution
by means of application of \Eq\eqref{J-Heun deformed} depending on an additional parameter
characterizing deformation degree. 
We discuss below an example of numerical implementation of the latter method.

\section{An example of realization of isomonodromic deformation of special double confluent Heun' equation}

\begin{figure}[h]
\scalebox{0.85}{
\includegraphics{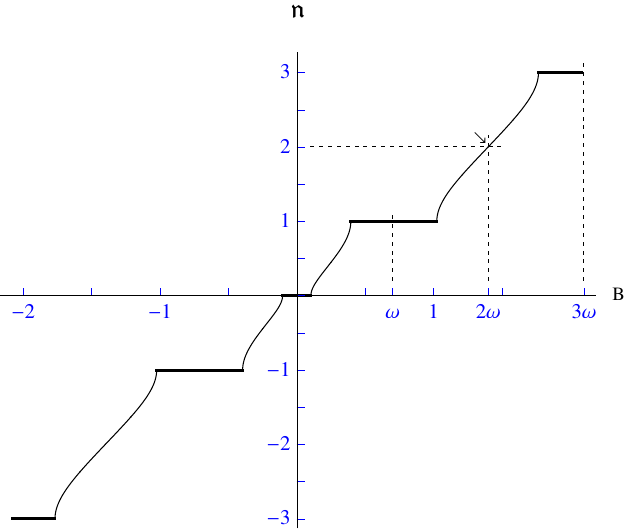}
}
\caption{ 
\small
IV-curve  similar to the one displayed in \Fig\ref{fig1} and plotted for the same
$\omega=0.7$ but 
$A=A_1=3.85399238\!\dots$ The place of the ``ghostly" Shapiro step is indicated by arrow.
}\label{fig6}
\end{figure}

First of all,
it is necessary
to chose some
phase-lock subarea and one of its constrictions to be further used as a starting data for subsequent computation.
There is no reason why
the case $\omega=0.7$ touched on in the above introduction for which a part of the corresponding geometrical structure
is depicted in \Fig\ref{fig3}
can not be used.
Several constrictions are therein present and any of them may be selected.
For definiteness, let us take $\ell=\ell_1=2$ and get the constriction with minimal (positive) level.
In \Fig\ref{fig3}, the section with constant $A$ passing through this point is shown.
The rotation number of the phase-lock subarea skewered on the vertical line  $\ell=2$  is equal to 2.
For such a value of $\ell$ the minimal level
the table~\ref{tab1} proposes
is
\begin{equation}\label{A_1}
A=A_1= 3.85399\!\dots
\end{equation}
One of
the  effects of this particular selection is illustrated by inspection of particularity of the corresponding version of the dependence
shown in \Fig\ref{fig1}.
Its graph for the above value of $A=A_1$
is displayed  in \Fig\ref{fig6}.

 It is seen that, as compared with \Fig\ref{fig1},
there is now no flat segment (Shapiro step) of the height $2=\rotnum$ (as well as for the height $-2$),
the graph being apparently smooth thereat. This has been expected and
one might say that it takes place simply because in the course of variation of $A$ towards $A_1$
the width of
the Shapiro step present in \Fig\ref{fig1} vanishes just at its value \eqref{A_1}.

It is worth reminding that
in accordance with \Eqs\eqref{AB vs lamu}
the choice \eqref{A_1} is
equivalent to the value of
$$\mu=\mu_1=2.75285\!\dots$$
originally obtained as solution to equation $\xi_2(\mu|0,7)=0$, see \Tab\ref{tab1}.

The triplet of  parameters $\ell_1,\mu_1,\lambda_1$, where
$$
\lambda_1=(2\omega)^{-2} - \mu_1^2=-7.06799\!\dots$$
(see \Eq\eqref{constconstr})
specifies the instance of \Eq\eqref{sDCHE} possessing solution holomorphic at zero.
We intend to apply to it the isomonodromic deformation described by \Eq\eqref{J-Heun deformed}.
The subject of deformation is however not \Eq\eqref{sDCHE} itself but the equivalent equation \eqref{J-Heun}. It includes the same parameter
$\ell=\ell_1$ whereas the other constant parameters have to be determined by means of the transformation  \eqref{parms generating}.
Accordingly, we introduce
\begin{equation} \label{starting data}
\troot_1=\mu_1^2=7.57819\!\dots,
\;
\bfupmu_1= - \lambda_1/\troot_1
=0.932675\!\dots
\end{equation}
If $\akat=\troot_1$ and $ \bfupmu=\bfupmu_1$ then \Eq\eqref{J-Heun} is equivalent to instance of \Eq\eqref{sDCHE} possessing
 solution holomorphic an zero. The same obviously concerns the current instance of \Eq\eqref{J-Heun}.

On the other hand, \Eq\eqref{J-Heun} is a particular case of \Eq\eqref{J-Heun deformed} arising when one of its parameters
vanishes, namely, when $\bfuplam =0$, whereas other eponymous parameters coincide.

The parameters  $\bfuplam,\bfupmu,\akat$ characterizing (in conjunction with $\ell$) the equation \eqref{J-Heun deformed}
are named constant parameters or simply constants but this only means their independence from the free complex-valued variable $\akaz$.
When considered unrelated to the latter, they may also be regarded as continuously varied themselves in some way  producing a continuum
of instances of equations \Eq\eqref{J-Heun deformed} which can be referred to (and named) as a deformation of the starting one
which can be chosen coinciding with \Eq\eqref{J-Heun}.
Among them, there is a deformation which is s known as possessing the property of  isomonodromicity \cite{JMU}
which in our case
reduces to existence
for each of the deformed equation
a  solution holomorphic at zero,
the point which is singular for all the linear second order equations we consider.
Such a deformation is constructed as follows.

Let us consider the quantities  $\bfuplam,\bfupmu$  as smooth functions of the parameter $\akat$ playing here role of a free real-valued variable
which is unrelated to the variable $t$ denoting physical time (see \Eq\eqref{RSJ or RCSJ})
 as well as to its dimensionless counterpart $\akatau$ (\Eq\eqref{overdamped RSJ}).
Let these functions be such that
\begin{equation}
\bfuplam(\troot_1)=0,  \bfupmu(\troot_1)=\bfupmu_1 \quad\mbox{(see \Eqs\eqref{starting data})}
\end{equation}
and let they obey \Eqs\eqref{eq for bfuplam}, \eqref{eq for bfupmu} on some segment of real around $\troot_1$.
With such $\bfuplam=\bfuplam(\akat),\bfupmu=\bfupmu(\akat) $
the set of equations \eqref{J-Heun deformed} arising in the course of variation of $\akat$ realises an isomonodromic deformation of \Eq\eqref{J-Heun}.

If it turns out that there is another root $\akat_2\not=\troot_1 $ of $ \bfuplam(\akat) $ then
 with such $\akat=\akat_2$ , $ \bfuplam=0$, $ \bfupmu=\bfupmu(\akat_2) $
\Eq\eqref{J-Heun deformed}
proves to be of the form of \Eq\eqref{J-Heun} and hence equivalent to some instance of \Eq\eqref{sDCHE}.
By construction, all these equations admit solutions holomorphic at zero.
Concerning \Eq\eqref{sDCHE}, this is what we search for.

On its own,
the above function $\bfuplam$ obeys \Eq\eqref{sPIII'} which is a particular case of   Third \Pain{} equations.
The equation is not a generic instance of the latter and, besides,
the value of $\ell$
is here constrained
to non-negative integers.
Nevertheless, our $\bfuplam$ proves to be a nontrivial solution of the noted equation
and hence should be qualified as a Third \Pain{} transcendent.
For brevity, the shorter name \PIIItran will be here used.
The most reasonable way of  determination of $\bfuplam(\akat)$ is here the using of numerical integration.

\begin{figure}[H]
\scalebox{1.10}{
\includegraphics{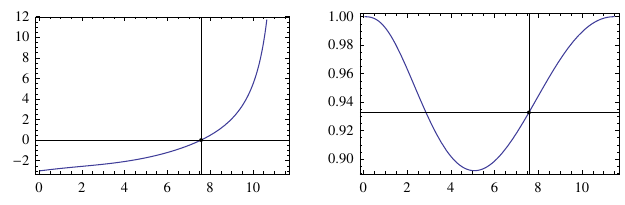}
}
\caption{
\small
 The graphs of numerical solutions $\bfuplam(\akat)$ (the left panel) and
   $\bfupmu(\akat)$ (the right one) to \Eqs\eqref{eq for bfuplam}, \eqref{eq for bfupmu}.
   The values of  the argument $\akat$ are plotted against the horizontal coordinate axes.
   The pairs of coordinate axes intersect at the points of specification of the initial data
   for the Cauchy problem the noted functions solve.
   It holds
    $\akat=\troot_1=7.57819\!\dots$ and $\bfuplam(\troot_1)=0$,
    $\bfupmu(\troot_1)=0.932675\!\dots$ thereat.
   Numerical integration providing the plotted  functions crashed on the both left (for $\akat$ slightly greater zero)
   and right (on the verge of $\akat\approx 11.4777 $) ends of the compounded domain.
   The value $\bfuplam$ exhibit thereat on the right is
   $\simeq 10^{12}$  (far beyond the range of the figure panel) whereas
    $\bfupmu$ is thereat indistinguishable from $ 1 $.
}
\label{fig7}
\end{figure}

The graphs of the functions $\bfuplam,\bfupmu$ meeting the above prescription are shown in \Fig\ref{fig7}.
\noindent
It is seen that $\bfuplam(\akat)$  is a monotonic function possessing the only zero (forced by a stipulation comprised in the initial data at the starting point of integration)
in the available domain. There is also strong evidence of its unboundedness when approaching to some point to the left of $\akatau=11.5$\:
Since $\bfuplam(\akat)$ belongs to  Third \Pain{} transcendents this point has to be its simple pole.
It also holds $\dbfuplam( \troot_1)=1$ (exactly) at the noted $ \bfuplam $ root.

Our current goal is to gain an another real root of $\bfuplam$ regarded as an analytic function.
We should obviously be sought on its distinct branch
which might be constructed by means of
analytic continuation across the (so far hypothetical) pole.

The efficient way of the handling of the issue in question is the passage from transcendents $ \bfuplam$ and $\bfupmu $
to  `mirrored' functions $ \dualam, \duamu$
by means of transformations \eqref{forward}.
The formulas are explicit and we may consider the latter functions as
already established on the segment
with the left boundary close to zero
and the right one somewhere to the right of $\troot_1$, for example, for $\akat<11.477 $.
It has to be noted that the very point $\akat=\troot_1$
is obviously the pole for the function $ \dualam(\akat)$ where it remains undefined.
Besides, as opposed to $\bfuplam$,
$ \dualam$
is regular on the right boundary of current temporary domain.
This can be confirmed by means of numerical integrating
of the equations \eqref{eqs for dualammu} which the functions $ \dualam,
\duamu$ have to obey on a segment extending more to the right.
Any point to the right of $\troot_1$  where values of $ \bfuplam$ and $\bfupmu $ are known can be used as the place of
specification of the initial data for the related Cauchy problem.
The result of solution of some of them is shown in \Fig 8. 
\begin{figure}[h]
\scalebox{1.20}{
\includegraphics{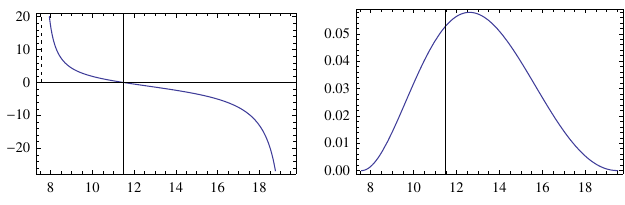}
}
\caption{
\small
The graphs of solution to equations \eqref{eqs for dualammu} begotten by solution to
equations \eqref{eq for bfuplam}, \eqref{eq for bfupmu}  imaged in \Fig\ref{fig7}.
The variable $\akat$ is plotted along the horizontal coordinate axes.
The left panel contains the graph of the function $\dualam(\akat)$; $ \duamu(\akat)$ is plotted on the right.
The numerical integrating
computing these functions 
crashes near $\akat=19.50603$.
The left end  $\akat=\troot_1$ of the available domain is, by construction, the position of the simple pole of $\dualam$.
The vertical dashed line in the left panel indicates the corresponding
shift and serves $\dualam$ asymptote.
The vertical coordinate axes are situated with the
shift $\tpole_1=11.47771\dots$ to pass in the left panel through the root
of $\dualam$ with value arising as a byproduct of the numerical integration.
$ \duamu$ vanishes at the domain edges.
}
\label{fig8}
\end{figure}

The numerical integration of \Eqs\eqref{eqs for dualammu} going
to the right as far as possible
crashes near $\akat=19.50603$\;
The left end of the arising domain is $\akat=\troot_1$ where,
by construction, the simple pole of $\dualam$ is situated.
Its observable root $\tpole_1=11.4777 \dots$
represents the  actual position  of the pole of $
\bfuplam(\akat)$
earlier predicted to be just of the above value but
estimated with uncertain accuracy.

As opposed to the increasing function $\bfuplam$  from \Fig\ref{fig7},
the function $\dualam$  from \Fig 8
is monotonically decreasing. Its derivative is negative and hence
it has to hold $\ddualam(\tpole_1)=-1$ at the root found.

\begin{figure}[h]
\scalebox{1.20}{
\includegraphics{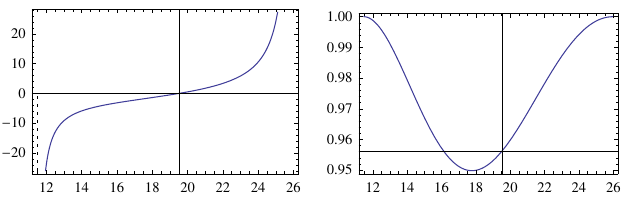}
}
 \caption{
\small
The graphs of solution to equations  \eqref{eq for bfuplam}, \eqref{eq for bfupmu}
 begotten by solution to
equations \eqref{eqs for dualammu}
imaged in \Fig 8. 
The variable $\akat$ is plotted along the horizontal coordinate axes.
The left panel contains the graph of the function $\bfuplam(\akat)$; $ \bfupmu(\akat)$ is plotted on the right.
The numerical integration
computing these functions
crashes near $\akat=25.98$.
The left end  $\akat=\tpole_1=11.4777\dots$ of the available domain
(the computed root of $\dualam$)
is, by construction, the position of the simple pole of $\bfuplam$.
The vertical dashed line in the left panel indicates the corresponding
shift and serves $\bfuplam$ asymptote.
The vertical coordinate axes are situated with the
shift $\troot_2=19.50603\dots$ to pass in the left panel through the root
of $\bfuplam$ arising as a byproduct of numerical integration.
Both left and right boundary values of $\bfupmu  $ are
indistinguishable from 1.
 }
\label{fig9}
\end{figure}

Now we are able to carry out the inverse transition from functions
$\dualam, \duamu $
to functions $\bfuplam, \bfupmu$   obeying \Eqs\eqref{eq for bfuplam} \eqref{eq for bfupmu} by means of the invertible transformation
\eqref{backward}
In view of explicitness of the latter,   $\bfuplam(\akat), \bfupmu(\akat)$
can be considered already established on the segment $(\troot_1,19.5]$.
There is however a pole situated at $\akat=\tpole_1$ and the
connected subdomain of continuity of $\bfuplam$ is terminated at this
boundary point. At the same time it is not terminated
by the right boundary point of the available domain
of $\dualam$.
A natural (i.e.\ analytic) continuation is simulated by means of numerical
integrating of \Eqs\eqref{eq for bfuplam} \eqref{eq for bfupmu}.
Its result is displayed in \Fig\ref{fig9}.

\noindent
The functions plotted therein represent analytic continuations of
eponymous ones plotted in \Fig\ref{fig7}. From them, the
new root $\troot_2=19.50603\dots$ of the function  $\bfuplam$ which is the data of our primary interest is determined.
As opposed to $\troot_1$, it was not set ``by hand'' via adapted initial data for a Cauchy problem and its existence had not been guarantied.
In this case it is gained  as an outcome of a nontrivial computation.

\begin{figure}[h]
\scalebox{1.20}{
\includegraphics{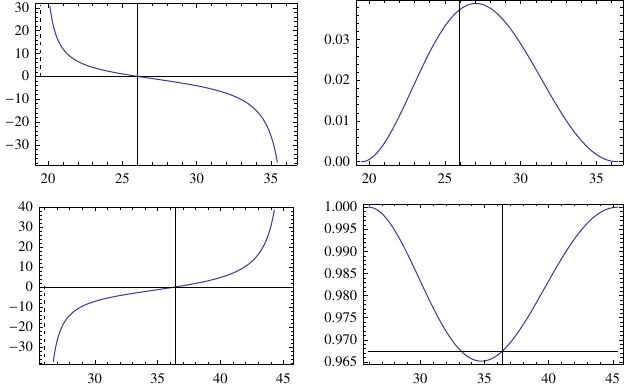}
}
 \caption{
\small
Graphs of pairs of functions $\dualam(\akat),\duamu(\akat)$ (shown in top-left and top-right panels, respectively)
and   $\bfuplam(\akat), \bfupmu(\akat)$  (in lower-left and lower-right ones, respectively) which verify
the systems of equations
 \eqref{eqs for dualammu}
and
\eqref{eq for bfuplam} \& \eqref{eq for bfupmu}, respectively.
The functions from the top panels
are defined on the segment $(\troot_2, \troot_3)=(19.5060\dots, 36.3976\dots)$,
where $\troot_3$ represents another root of the function $\bfuplam(\akat) $ from the lower-left panel arising
as a byproduct of integration.
There are asymptotes for the function $\dualam$ with these shifts.
The domains of functions in the lower panels are the segment
$(\tpole_2, 45.39)$, where $ \tpole_2= 25.9806\dots$ is another root of the function $\dualam$ from the top-left panel
arising as a byproduct of integration.
In the subdomain $(\troot_2,\tpole_2)$ the functions  $\dualam(\akat),\duamu(\akat)$
are connected with
the functions  $\bfuplam, \bfupmu$  shown in Fig 9 by transformations \eqref{forward},\eqref{backward}.
In the adjacent subdomain  $(\tpole_2,\troot_3)$  the above functions from the top panels are connected with functions
 $\bfuplam, \bfupmu$ from the lower panel by the same transformation. At boundaries of the corresponding domains, the functions
$\duamu$ and $ \bfupmu $ are indistinguishable from 0 and 1, respectively.
 }
\label{fig10}
\end{figure}

The procedure outlined above can be repeated using the second root $\troot_2$ of $\bfuplam$ in place of its first root  $\troot_1$.
In this way nothing fundamentally new may (and do) arise and it makes no sense to provide here its details.
The relevant final data in graphical form are to be displayed instead.
They are given in \Fig \ref{fig10} (page \pageref{fig10}).

These graphs look qualitatively similar to the corresponding ones displayed above. The most useful result they gain is the new, third, root $\troot_3$
of the function $\bfuplam(\akat)$ plotted in the left-lower panel. It is also important to have the value of $\bfupmu(\troot_3)$ --- as well as
 $\bfupmu(\troot_2)$ which can be obtained from the data provided by the function plotted in the right panel of \Fig \ref{fig9}.

It has to be kept in mined that the graphs labeled by eponymous functions shown above in different figures represent the
fragments of a graph of a single analytic function evaluated on certain adjacent finite segments of the real axis.
These segments could be stacked together in a straightforward way
giving rise to a unified picture for each of the four functions $\bfuplam, \bfupmu, \dualam, \duamu$
we have considered.

Of them, $\bfuplam$ and $\dualam $ (both representing \PIIItran{}s) reveal both zeros (at which the slopes of graphs are equal to $\mp1$) and
poles (of known residues).

On the contrary, the functions $\bfupmu, \duamu$ are continuous and, then, everywhere smooth
 --- or, more exactly,  holomorphic.
Besides, in the domain we consider,
their values are enclosed within the interval $[0,1]$ osculating, as graphs,
one of its
either top ($ \bfupmu $) or bottom ($ \duamu$) boundary lines.

Such a heuristic observation hints at reasonability of inspection of behavior of subsidiary expressions
$(1- \bfupmu(\akat))^{1/2}$ and $ \duamu(\akat)^{1/2}$. The corresponding graphs plotted upon the combined domain
are shown in \Fig 11.
\begin{figure}[h]
\scalebox{1.15}{
\includegraphics{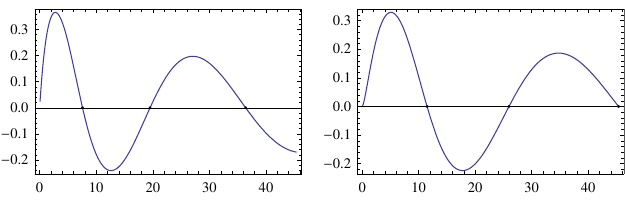}
}
\caption{
\small
The graphs of quantities 
$ \duamu(\akat)^{1/2}$
(the left panel)
and
$(1- \bfupmu(\akat))^{1/2}$
(the right one).
With minor exception they reincarnate a part of the data already exhibited
in the right panels of figures \ref{fig7}--\ref{fig10}.
 }
\label{fig11}
\end{figure}
The roots of the left-panel function are equal to $\troot_1, \troot_2,\troot_3$. They thus coincide with roots of $ \bfuplam $.
The roots of the function on the right are equal to $\tpole_1, \tpole_2,$ and $\tpole_3=45.3910\!\dots$
These provide positions of poles of the
same function. Since $\bfuplam (\akat)\dualam(\akat)\equiv \akat$,
the roles of the above roots and poles
in relation to the function $\dualam$
is provided by 
interchanges `is a root'\:$\leftrightharpoons$\:`is a pole'.

The apparent smoothness of functions plotted in \Fig\ref{fig11} and importance of roles of their roots clearly indicate that they deserve a separate consideration.


\section{Some more formulas}\label{sec3}

The following statement takes place.
\begin{statement}\it
If the functions $\bfupmu(\akat)$ and $\duamu(\akat)$ obey the equations  \eqref{sPIII' for bfupmu}  and \eqref{eq for duamu}, respectively,
then the
two instances of the function $\akav(\akat)$ and the constant $M$ defined as follows
\begin{equation}\label{these are bounded}
\hspace{-0em}
\begin{aligned}
&  \duamu(\akat)^{1/2}
\\[-2.5ex]
 \mbox{\raisebox{2.0ex}{\hspace{-2.0ex}\llap{$\rightharpoondown\hspace{0.55ex}$} }} &
\llap{$\akav(\akat)  =\hspace{-0.06em}\Big\langle$ \hspace{1.1em}}
\\[-2.5ex]
 \mbox{\raisebox{-0.1ex}{\hspace{-0.1ex}\llap{$\rightharpoonup\hspace{0.6ex}$} }}
&(1- \bfupmu(\akat))^{1/2}
\end{aligned}
\hspace{0.5em}\mbox{ in conjunction with }\hspace{4.3em}
\begin{aligned}
&  \llll-1
\\[-2.5ex]
 \mbox{\raisebox{2.0ex}{\hspace{-2.0ex}\llap{$\rightharpoondown\hspace{0.55ex}$} }} &
\llap{$M =\hspace{-0.06em}\Big\langle$ \hspace{1.1em}}
\\[-2.1ex]
 \mbox{\raisebox{-0.2ex}{\hspace{-2.0ex}\llap{$\rightharpoonup\hspace{0.6ex}$} }}
&\llll
\end{aligned} \hspace{-4em}
\end{equation}
obey corresponding instances of the equation
\begin{equation}\label{the second one}
0=
\ddv+
\bigg(\!\dakav^2 - \left({M\over 2 \akat}\right)^{\!\!2} \bigg) { \akav \over 1 - \akav^2}
+{\dakav \over \akat }
+{\akav \over \akat } (1-\akav^2)
\end{equation}
\end{statement}

The origins of functions $\duamu(\akat)$ and  $\bfupmu(\akat)$ concordant with the noted equations they obey
imply the following statement:
\begin{statement}\it
If $ \akat_\asteRisk>0 $ and $\bfuplam(\akat_\asteRisk)=0$ then the following particular values of the functions
 $\akav$ introduced in \eqref{these are bounded} and their derivatives $\dakav  $ are observed:
\begin{equation}\label{onshell}
\hspace{3em}
\begin{aligned}
&  0
\\[-2.5ex]
 \mbox{\raisebox{2.0ex}{\hspace{-2.0ex}\llap{$\rightharpoondown\hspace{0.55ex}$} }} &
\llap{$ \akav(\akat_\asteRisk)  =\hspace{-0.06em}\Big\langle$ \hspace{1.1em}}
\\[-2.5ex]
 \mbox{\raisebox{0.1ex}{\hspace{-0.1ex}\llap{$\rightharpoonup\hspace{0.5ex}$} }}
&(2\omega_\asteRisk\,\akat_\asteRisk^{1/2})^{-1}
\end{aligned},
%
\hspace{6em}
\begin{aligned}
&  -(2\omega_\asteRisk\,\akat_\asteRisk)^{-1}
\\[-2.5ex]
 \mbox{\raisebox{2.0ex}{\hspace{-2.0ex}\llap{$\rightharpoondown\hspace{0.55ex}$} }} &
\llap{$ \dakav(\akat_\asteRisk)  =\hspace{-0.06em}\Big\langle$ \hspace{1.1em}}
\\[-2.5ex]
 \mbox{\raisebox{0.01ex}{\hspace{-0.5ex}\llap{$\rightharpoonup\hspace{0.6ex}$} }}
&-3(4\omega_\asteRisk\,\akat_\asteRisk^{3/2})^{-1}
\end{aligned}
\end{equation}
\end{statement}
It follows from the above relations that
for the lower instance of $\akav$
\begin{equation} \label{islowing}
\omega_\asteRisk=
\sqrt{ -\dakav(\akat_\asteRisk)
\over
6\akav(\akat_\asteRisk)^3
}
\end{equation}

\section{Effectual way of transportation of constrictions of phase-lock area}

\begin{figure}[H]
\scalebox{1.15}{
\includegraphics{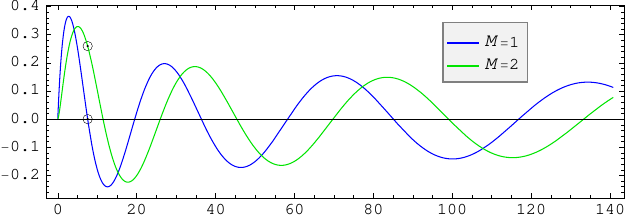}
}
\caption{
\small
Graphs of solutions to the equation \eqref{islowing} with indicated values of the parameter $M$.
The variable $\akat$ is plotted along the horizontal coordinate axis.
The initial data for the Cauchy problems the imaged functions solve are set at centers of small circles at which  $\akat=\troot_1=7.57819\!\dots$
Specifically, in accordance with \Eqs\eqref{onshell} (with $\akat_\asteRisk=\troot_1,\;\omega_\asteRisk=\omega=0.7$) it is assigned
$\{\akav,\dakav\!\}(\troot_1)=\{0,$  $-0.0942554\dots \} $
and
$\{\akav,\dakav\!\}(\troot_1)=\{0.259471\!\dots,$  $-0.0513588\dots\} $
for upper and lower cases, respectively.
The left part of the graph with $M=1$ coincides with the graph in the left panel of \Fig\ref{fig11}.
The graph in its right panel is identical to the left part of the above graph corresponding to \mbox{$M=2$}.
 }
\label{fig12}%
\end{figure}

It seems natural to inspect computational capability of application of \Eq\eqref{the second one} to the problem of determination
of roots of \PIIItran{} $ \bfuplam $ which serves a parameter of \Eq\eqref{J-Heun deformed}
describing isomonodromic deformation of \Eq\eqref{J-Heun} equivalent to \Eq\eqref{sDCHE}.
We will solve both its instances, one with $M=\llll-1=1$ and another with $M=\llll=2$,
setting the corresponding Cauchy problems.
The initial value of the free real variable $\akat$ is the $ \bfuplam $ root $\troot_1$ assumed to be given
(in fact, it had been obtained from the minimal positive root of the function $\xi_2(\mu|\omega)$,
see \Eqs\eqref{withnotherarg}, \eqref{def of xi}, \eqref{R-matrix}, \eqref{M-matrix}, \eqref{parms generating}.
and \Fig\ref{fig5}). The necessary initial data is provided by \Eqs\eqref{onshell}
in which $\akat_\asteRisk=\troot_1=7.57819\dots$ and $\omega_\asteRisk=\omega=0.7$.
Computations are straightforward since it everywhere holds $|\akav|<1$ and `the right hand sides' of the equations integrated are well defined and regular for all values of $\akat$,
$\akav$, and $\dakav$ which occur. The only exception is the point $\akat=0$ and, indeed, the code\footnote{{\sc mathematica}'s routine {\tt NDSolve}}
crashes slightly to the right of it. The solutions thus obtained are shown in \Fig\ref{fig12},
\noindent
They represent apparently damped oscillations around zero with interspersing roots.
Their left halves coincides with graphs shown in \Fig\ref{fig11}. Now, however, their domain is twice greater --- and could be easily extended
further to the right. Accordingly, more roots we are search for are exhibited allowing to conjecture that their total set is countable.
Specifically,
the roots of the solution  to \Eq\eqref{the second one}
with $M=1$
(equal to $\duamu(\akat)^{1/2}(\akat)$, see \Eq\eqref{these are bounded}). are listed in table \ref{tab2} (page \pageref{tab2}).
\noindent
\begin{table}[t]\small
\begin{tabular}{cllllll}
\hline
$\asteRisk$
& \multicolumn{1}{|c|}{$\troot_\asteRisk$}
& \multicolumn{1}{|c|}{$\omega_\asteRisk$}
& \multicolumn{1}{|c|}{$\mu_\asteRisk$}
& \multicolumn{1}{|c|}{$\lambda_\asteRisk$}
 &
\multicolumn{1}{|c|}{$A_\asteRisk$}
 &
 \multicolumn{1}{|c}{$B_\asteRisk$}
\\
\hline\\[-1.8ex]
1&\small 7.57819\dots&\small0.7&\small 2.75285\dots&\small -7.06799\dots&\small 3.85399\dots&\small1.4 \\
2&\small19.5060\dots&\small0.540363\dots&\small4.41656\dots&\small-18.6498\dots&\small 4.77309\dots&\small1.080725\dots\\
3&\small36.3976\dots&\small0.459258\dots&\small6.03304\dots&\small-35.2123\dots&\small5.54145\dots&\small0.918516\dots\\
4&\small 36.3976\dots&\small 0.407100\dots&\small7.63263\dots&\small-56.7485\dots&\small6.21449\dots&\small0.814201\dots\\
5&\small85.0801\dots&\small0.369732\dots&\small9.22389\dots&\small-83.2513\dots&\small 6.82074\dots&\small0.739464\dots\\
6&\small116.863\dots&\small 0.341196\dots&\small10.8103\dots&\small-114.716\dots&\small 7.37687\dots&\small0.682391\dots\\
\multicolumn{1}{c}{$\cdots$}&\multicolumn{1}{c}{$\cdots$}&\multicolumn{1}{c}{$\cdots$}&\multicolumn{1}{c}{$\cdots$}
&\multicolumn{1}{c}{$\cdots$}&\multicolumn{1}{c}{$\cdots$}&\multicolumn{1}{c}{$\cdots$}\\[-0.1em]
\hline
\\[-1.8ex]
\end{tabular}
\caption{\small 
Values of roots $\troot_\asteRisk$ of the function $\akav(\akat)$
obtained by means of numerical integration of \Eq\eqref{the second one} with $M=1$
whose graph is shown in \Fig\ref{fig12} and derived quantities
serving constant parameters of \Eq\eqref{sDCHE} (with $\llll=2$) and \Eq\eqref{overdamped RSJ}
which provide the points
of triviality of the monodromy group for the former equation
and the points
of constrictions of phase-lock area for the latter one.
}\label{tab2}
\end{table}
Besides the roots, their important derivatives are given as well.

In particular, the third column contains the values of dimensionless frequencies $\omega$
entering the basic equations \eqref{overdamped RSJ}, \eqref{sDCHE}. It is determined by the formula \eqref{islowing}
and thus requires additional computation of solution to \Eq\eqref{the second one} with $M=2$ (that has prudently been carried out).

The fourth column  contains the values of the parameter $\mu$ from \Eq\eqref{sDCHE}
determined in accordance with the second formula in \Eq\eqref{parms generating}.
Parameters $\lambda$ from the same \Eq\eqref{sDCHE} tabulated in the fifth column are determined from the preceding two ones
in accordance with the second of equalities \eqref{constconstr}.
The last but one and the last columns list values of the parameters $A,B$ also entering  \Eq\eqref{overdamped RSJ}.
The corresponding formulas are comprised in the system of \Eqs\eqref{AB vs lamu}.

Now we have several packages of data for equations \eqref{overdamped RSJ} and \eqref{sDCHE} expected to provide them
with outstanding properties. Namely, the latter equation   with any instance of triplets of its parameters contained in
\Tab\ref{tab2}
has to admit a solution holomorphic at zero and being therefore an entire function. Other solutions may be singular thereat but all they are holomorphic
 (and single valued) in
$\mathbb{C}\backslash0$. In other words, no branching over zero is observed.
This means that the monodromy groups of these equations which all are singular at zero (and at infinity) are trivial.
Such a triviality is in our case the property that is conserved in the process of isomonodromic deformation
referred to above.

As to \Eq\eqref{overdamped RSJ}, the  triplets of its constant parameters listed in \Tab\ref{tab2}
represent coordinates of the points of constrictions of phase-lock areas
(or, taking into account incorporation into consideration of variation of the frequency $\omega$, the lines) in its parameter space.
We have seen the picture of the flow of integral lines of \Eq\eqref{overdamped RSJ} with parameters from a constriction
point in \Fig\ref{fig3}.  These parameters are also contained in \Tab\ref{tab2} occupying relevant positions in its first line.
A two-fold periodicity is characteristic for such an evolution and it is revealed by all solutions of the equation.

\begin{figure}[H]
\scalebox{1.15}{
\includegraphics{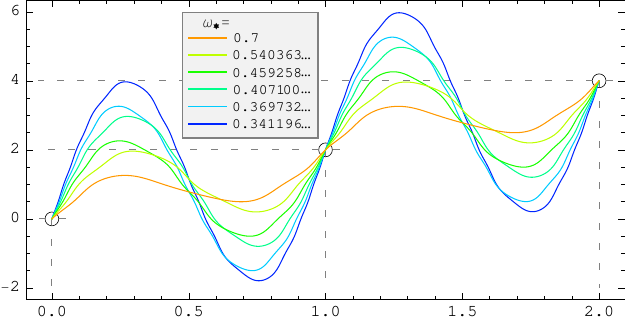}
}
\caption{  
   Graphs of solutions   $\phi=\phi_\asteRisk$, $\bullet=1,\dots6,$ to \Eq\eqref{overdamped RSJ}
   started at zero.
   Constant parameters  $A,B,\omega$ for the latter
   are taken from the table \ref{tab2}. The quantities
   $(2\pi)^{-1}\omega_\asteRisk \akatau$ are plotted along the horizontal coordinate axis, the values of
   $(2\pi)^{-1}\phi_\asteRisk(\akatau)$  are plotted along the vertical one.
   The points of everyone's intersections
   whose coordinates prove to be integer
   are indicated by small circles centered thereat.
Ineluctable visual defects due to limited accuracy of computations are here imperceptible
   and all these somewhat wobbling curves are truly such as they look.
}
\label{fig13}
\end{figure}

We will demonstrate the existence of such a periodicity of solutions for all six instances of  \Eq\eqref{overdamped RSJ}
inferred from the data contained in
\Tab\ref{tab2}.
However, we do not here engage `a continuum of available solutions' similar to the one sketched in \Fig\ref{fig3} .
Instead, only one of them for each selected equation is get.
For simplicity of a clear comparison, it is reasonable in our case to use solutions `started from zero'.
Their graphs are shown in \Fig\ref{fig13}.
The periods characteristic for solutions of different equations we consider depend on the frequency $\omega_\asteRisk$ and are not the same.
To make direct comparison feasible,
not very variable $\akat$ but
the results of its normalizations $(2\pi)^{-1}\omega_\asteRisk\akat$
are plotted along the horizontal axis.
Besides, for the sake of clarity, the solutions  $\phi_\asteRisk$ are also normalized, this time on the same level and 
values of
 $(2\pi)^{-1} \phi_\asteRisk $
are plotted along the vertical axis.

In the resulting figure all the imaged curves intersect at points with integer coordinates $(0,0)$, $(1,2)$, $(2,4)$ (and, obviously, so forth).
Thus the precise periodicity of all solutions
up to the same incrementing on each period by $2=\llll$ (equal to the rotation number $\rotnum$) is observed.

This property is possessed by all solutions to the equations under consideration.
The expected effect of isomonodromic deformations referred to above can thus be considered to be numerically confirmed.

\section{Discussion}

The geometric structure of phase-lock areas in the space of parameters of \Eq\eqref{overdamped RSJ}
applied to the modeling of the response of the state of overdamped Josephson function to a typical
external impact represented by the equation right hand side
is of notable interest from various points of view.
It has been shown that in case of moderate values of the corresponding parameters $A,B,\omega$
each section $\omega=\:$const$\;>0$ of their subset on which the state of asymptotic phase-lock is observed (2D phase-lock area)
can be imaged as a collection of non-intersecting garland-like  subareas comprised from somewhat deformed lemon-like open sets
pairwise concatenated as their corner points in such a way that for each subarea
their combined boundary proves to be constituted by two smooth oscillating curves with transversal intersections.
See \Fig\ref{fig3}
from which more complete picture can be obtained utilising its mirror-symmetry with respect to the both coordinate axes.

It is important to note that each phase-lock subarea is associated with certain rotation number $\rotnum$ which all are of integer values.
Specifically, it holds $\rotnum=0$ for all points of the phase-lock subarea around the symmetry axis $B=0$.
When passing further one by one through subsequent phase-lock subareas, the rotation number corresponding to  their points
gradually increments to the right and decrements to the left by the unit,
all the integers appearing in this way.

The states provided by isolated (when considered in sections $\omega$=const)
points of concatenation of nearby `lemons' constituting given phase-lock subarea, been formally outside the latter,
reveal the same rotation number.
These points are named constrictions except the ones situated on the horizontal coordinate axis $A=0$. The latter are named growth points.

All the constrictions of some phase-lock subarea are characterised by the the same horizontal shift (or bias) $B=\omega\,\rotnum$ while
it holds
$B=((\omega\,\rotnum)^2+1)^{1/2}$ for its growth point.
An exception is provided by the case $\rotnum=0$: there is no growth point for this phase-lock subarea at the
coordinate
center or somewhere else.
One may consider such point to be blown up and turned into the segment $[-1,1]$ whose open part belongs to the phase-lock subarea
and boundary points play no peculiar roles.
It is worth noting however that, as one concerns to physics, for the states corresponding to points from the above segment
there exists non-zero constant current through the Josephson junction
regardless of the fact that no difference of the electric potential is there observed.
At the segment boundary points the maxima of intensity of such a miraculous DC  are reached.

 The location of constriction points is definitely important, though not only, telling quantitative characteristic of the geometry of
phase-lock area. It can be inferred from diagrams similar to the one shown in \Fig\ref{fig3} which consolidate the data produced by numerous
numerical integration of the very equation \eqref{overdamped RSJ} but this way is definitely not the best one.
More efficient method is based on the search for real roots of the function $\xi_l$ defined by the explicit formula \eqref{def of xi}
or its modified version \eqref{withnotherarg}. The remarkable property of these roots   is the ensuring existence of
a solution to \Eq\eqref{sDCHE} holomorphic at zero and thus representable by an entire function.

The function $\xi_\llll$ remains mainly unexplored. Some exploratory computations show that
for non-negative integer $\llll$ and positive $\mu,\omega$
it reveals a weakly decaying smooth oscillations around zero.
Several such examples
can be seen in \Fig\ref{fig5}. The desired roots thus exist and can easily be obtained ---   numerically. Some of them are listed in
\Tab\ref{tab1}. The location of all the constrictions seen in \Fig\ref{fig3} is encoded in these data.

In the present work we focus on another method of determination of location of constrictions seeming perhaps somewhat more sophisticated than the aforementioned ones.
As a matter of fact, it does not compute the location of constriction from the very beginning but takes
data specifying some known constriction to (formally infinite) sequence of similar but distinct constriction datasets.

A noteworthy feature of the method is the non-constancy of the dimensional frequency $\omega>0$ in the results it generates.
For the sequence of the data characterizing constrictions it provides, the frequencies constitute an own sequence
which proves to be monotonously descending. Its meaning is currently incomprehensible.

The aforementioned computation is based on
the process of so called deformation of the special double confluent Heun equation \eqref{sDCHE} reducing formally
to the replacement of the equivalent equation \eqref{J-Heun}
with the linear second order equation \eqref{J-Heun deformed}. As opposed to \Eqs\eqref{sDCHE} and \eqref{J-Heun},
the latter equation involves the four rather than three parameters represented by the quantities $\llll, \akat, \bfuplam,  \bfupmu$
(among which $ \bfuplam,  \bfupmu $ are unrelated to the parameters $\lambda,\mu$ from \Eq\eqref{sDCHE})
which are considered to be independent of the free complex variable $\akaz$ (distinct of $z$ from \Eq\eqref{sDCHE}).
An additional `degree of freedom' thus appears at our disposal and
under appropriate conditions the corresponding variation carried out `by hand' can be regarded as deformation ---
a connected family of continuously modified equations containing the original one.
It remains to clarify what
is here being deformed and what the above conditions mean.

In conjunction with the free complex variable $\akaz$ the variable $\akat$ independent of it plays the basic role. Generally speaking, it
can be complex values but in frameworks of the problem we deal with it is admissible (and reasonable) to set $\akat\in\mathbb{R}_+$. The parameter $\llll$
is considered to be a fixed non-negative integer (as in all the numerical examples outlined before). The remaining quantities $\bfuplam, \bfupmu$ from
\Eq\eqref{J-Heun deformed} are no longer truly constants.
Instead, they are functions of $\akat$ of which is claimed to obey the system of the non-linear
first order differential equations \eqref{eq for bfuplam}, \eqref{eq for bfupmu}. Upon elimination of the unknown  $ \bfupmu $, a particular case \eqref{sPIII'}
of the \PainIII{} equation \eqref{eq. PIII'} arises. $\bfuplam$ represents therefore an instance of \PainIII{} transcendents (\PIIItran{}s).
Since \Eq\eqref{eq for bfuplam} is the linear algebraic equation with respect to $\bfupmu(\akat)$
and similar relation takes place for \Eq\eqref{eq for bfupmu}  and the function $\bfuplam(\akat)$
the transitions $\bfuplam\leftrightarrows\bfupmu$ are reversible.

Now we are able to determine what an object our `deformation' deforms. Indeed, let us assume there is such $\akat=\akat_1>0$ that $\bfuplam(\akat_1)=0$.
Then it can be shown that \Eq\eqref{J-Heun deformed} becomes coinciding with some instance of \Eq\eqref{J-Heun} which, in turn, is equivalent to
some instance of \Eq\eqref{sDCHE}. The constant parameters of the latter are determined by values of the quantities $\akat_1$ and $\bfupmu(\akat_1)$.

Conversely, given \Eq\eqref{overdamped RSJ} associated with some constriction of a phase-lock subarea we can get its constant parameters $A_1,B_1,\omega_1=\omega$,
compute $\lambda_1,\mu_1$ in accordance with \Eqs\eqref{AB vs lamu}, \eqref{constconstr}, and finally convert the latter to $\akat_1$ and $ \bfupmu(\akat_1) $
by means of \Eqs\eqref{parms generating}.

We are now in a position to be able to refer to existence of solution to Cauchy problem for \Eqs\eqref{eq for bfuplam}, \eqref{eq for bfupmu}
with initial data specified at $\akat=\akat_1$ which assumes in particular that
$\bfuplam(\akat_1)=0$. The curve $\{\akat,\bfuplam(\akat), \bfupmu(\akat)\}$ in the space of parameters of \Eq\eqref{J-Heun deformed}
(in which $\llll$ is assumed to be fixed) just defines a deformation
of 
a particular equation \eqref{J-Heun}
containing it as its `initial element' to be `deformed'.

Such a deformation would 
produce new instances of \Eq\eqref{J-Heun} provided $\bfuplam(\akat)$ has other non-zero roots in addition the root $\akat_1$
imposed in its very definition. 
Indeed, let $\akat_j>\akat_1$ be such root and let us get the value of $ \bfupmu(\akat_j)$. Then the
way inverse to the one which had above lead from $A_1,B_1,\omega_1$ to  $\akat_1$ and $ \bfupmu(\akat_1) $
allows one to obtain the parameters concretizing \Eq\eqref{J-Heun} which will obviously be in the same relation
to `the deformation' \eqref{J-Heun deformed} as the original \Eq\eqref{J-Heun}.
In particular,
the new frequency can be found from \Eq\eqref{om determination}.

The obvious principal point of our discussion still unexplained is the meaning of the assignment of \PainIII{} equation
to the role of determination of parameters of the deformation \eqref{J-Heun deformed}.
It turns out that such a solution makes the latter monodromy preserving.

The equation \eqref{J-Heun deformed} and its particular case \eqref{J-Heun} are singular at zero.
When resolved with respect to the higher derivative, their right hand sides involve coefficients diverging at zero with the order 2.
As a consequence, in general case they admit no solutions which are single valued in connected open neighborhoods of zero which is thus their branch point.
In other words, analytic continuation of a solution to one of the noted equation along any non-contractible loop enclosing the center of the complex plane
does not coincide, upon arriving at the starting point, with the original function. Since it is still a solution and the equation it obeys  is linear homogeneous,
it is possible to
represent the action of the noted analytic continuation along a loop as a linear automorphism on the space of all solutions.
The set of all such automorphisms constitute the group which is named monodromy group.
It can then be shown that
if the deformation we consider are specified by the data generated by solutions to the equations \eqref{eq for bfuplam}, \eqref{eq for bfupmu}
then the monodromy group is kept unchanged.
Moreover, restricting to appropriately chosen bases of the solution spaces of the deformed equations, the
very linear transformations equivalent to analytic continuations around zero prove to be also unchanged.

An additional issue has also to be touched upon for the sake of completeness. As opposed to \Eq\eqref{J-Heun}, `the right hand side' of the equation  of deformation
\eqref{J-Heun deformed}, besides the singularity at zero, suffers from the pole situated at $\akaz = \bfuplam$. Such a distinction could result in existence
of the additional generator of the corresponding monodromy group making the structure of monodromies of the initial and deformed equations incompatible.
It can be shown however that the singularity of \Eq\eqref{J-Heun deformed} observed at $\akaz = \bfuplam$ produces a rather weak impact to the
structure of solutions. In particular, all they prove to be single-valued in vicinity of the additional singularity.
Hence, concerning monodromy properties, the noted feature of \Eq\eqref{J-Heun deformed} does not play a notable role and may be ignored.

Fortunately,
the following
the full-fledged notion of monodromy is excessive for frameworks of the problem we discuss.
All what we actually need here is the case of a null monodromy, i.e.\ the trivial monodromy group,
meaning that all solutions to \Eq\eqref{J-Heun} (and then, due to isomonodromicity of the deformation, to \Eq\eqref{J-Heun deformed})
are single-valued near zero --- although some of them still diverge thereat. This is just what the cases of parameters corresponding to
constriction points on phase-lock subareas assume.

Indeed, for such parameters \Eq\eqref{sDCHE} admits a non-trivial solution $E(z)$ holomorphic at zero.
Another solution $(\varhash E)(z)$ which is linear independent with  $E$ is obtained by means of the transformation \eqref{diez}.
It is important to note that since $\llll\ge0$ is integer this second solution
is singular at zero but still
single-valued. The pair $\{E, \varhash E\}$ constitute the basis
of the space of solutions
to \Eq\eqref{sDCHE} which all also possess therefore the latter property.

Now we have everything necessary for implementation of the approach outlined.
More or less arbitrarily, we chose to carry out transportation of the
constriction point
of minimal level 1 and non-minimal rotation number 2. Its position is indicated in \Fig\ref{fig3} as intersection of the dashed vertical line shifted by
$2\omega$ and horizontal dashed line $A=3.85399\!\dots$, the latter number being found in an independent way, namely, by means of numerical solving of
\Eq\eqref{perfectness}, see also \Fig\ref{fig5} for recognition of the related root (value of $\mu$)  $2.75\!\dots$

The first step assumes numerical integration of \Eqs\eqref{eq for bfuplam}, \eqref{eq for bfupmu} extending initial data
specified at $\akat=\akat_1=\mu^2$ (see the second equation from \Eqs\eqref{parms generating})
and establishing $\bfuplam(\akat_1)=0,\bfupmu(\akat_1)= -\lambda/\akat_1$ (the first one from \Eqs\eqref{parms generating}) as far as possible.

The result shown in \Fig\ref{fig7} seems to be a bit disappointing.
The domain of the solution proves to be bounded from the both sides
by crashes of the code runs and the \PIIItran{} $ \bfuplam(\akat) $ thus obtained is monotonously growing. It has therefore  no roots
distinct from the one emplaced at the
point of specification of the initial data,
The following observation seeming however to be worth mentioning: the function $ \bfuplam(\akat) $ is perhaps
smooth at the point $\akat=0$,
where the equation \eqref{sPIII'}  it obeys is singular,
manifesting thereat negative value and positive derivative.

Near the right end of the available domain,  $ \bfuplam(\akat) $ is apparently unbounded.
Such an obstacle to our computational procedure should not be insurmountable.
Its appearance  simply means that there exists a pole of the \PainIII{} transcendent $\bfuplam$ extended to a complex domain which
are allowed to have them by the theory (in fact, definition) of \Pain{} transcendents  and this pole (necessarily simple)
proves to be  situated on the real axes
just behind the right end of the domain we have obtained. 
Accordingly, the  situations could be resolved by means of analytic continuation `around the rightful singularity'.
This would require coming out from real axis to the complex plane.

There is however another more elegant way of the problem handling based on a discrete symmetry  of the \PIIIeq{} under consideration.
Namely, the involutive
transformation described by \Eqs\eqref{forward}, \eqref{backward} converts
\Eqs\eqref{eq for bfuplam}, \eqref{eq for bfupmu}
to
\Eqs\eqref{eqs for dualammu} associated with \Eqs\eqref{sduaPIII'} also representing a particular case of \PainIII{} equations.

The form of interdependence
of $\bfuplam$
and 
$\dualam $ says that the locations of poles and roots of the former coincide with the ones of locations, respectively, of roots and poles of the latter.
Thus $\akat_1$ is the pole of $\bfuplam(\akat)$ which is then well defined on the right of it at least within the available part of domain of $\bfuplam$
which is free of other roots.
Accordingly,
let us choose any point $\tilde\akat$  therein  and compute
$\dualam(\tilde\akat),  \duamu(\tilde\akat)$ in accordance with \Eqs\eqref{forward}.
Then we can set and numerically solve the Cauchy problem for
\Eqs\eqref{eqs for dualammu} on a domain as long as possible.
As one should expect, the latter proves to be limited from the both sides that is seen in \Fig\ref{fig8} displaying the corresponding solution.

The \PIIItran{} $ \dualam $ is also monotonic  but, as opposed to $ \bfuplam $, it is decreasing.
We had started its determination at a place near which $ \bfuplam $ is regular and non-zero.
$ \dualam $ is here also regular but it is regular at and somewhat beyond the location of apparent divergence of $ \bfuplam $ as well.
 This `critical point'
is merely a root of $ \dualam $  playing at the same time the role of a pole of $\bfuplam$, see the comments to \Fig\ref{fig8}.
Hence its notation $\tpole_1$.

Next, let some $\tilde{\tilde\akat}$ be chosen a little to the right of $\tpole_1$.
We know the values of $ \dualam(\tilde{\tilde\akat})$ and $\duamu(\tilde{\tilde\akat}) $
and can compute $\bfuplam(\tilde{\tilde\akat}),  \bfupmu(\tilde{\tilde\akat})$ by the formulas \eqref{backward}.
Then we solve the Cauchy problem for \Eqs\eqref{eq for bfuplam}, \eqref{eq for bfupmu} using these data as the initial one.
The result is shown in \Fig\ref{fig9}.

The function whose graph is there displayed in the left panel is the analytic continuation of the function with graph displayed in the left panel in
\Fig\eqref{fig7}. In other words they represent the values of a single analytic function  which has been denoted $\bfuplam $
 on a segment of real axis with a point  removed.

The same relation binds the functions with graphs displayed in the right panels with the difference that
the domain of such a combination is the full union of segments shown in each of the figures in question.

These relations show that \PIIItran{} $\bfuplam$ 
has another root $\troot_2 $ whose location has been determined up to a known accuracy
secured in the process of numerical integration of differential equations.

Its value and the value of  $ \bfupmu( \troot_2 ) $ provide  the ones of the constant parameters $\lambda=\lambda_2, \mu=\mu_2, \omega=\omega_2$
(see \Tab\ref{tab2} for their numerical values)
in accordance with \Eqs\eqref{parms generating}, \eqref{om determination}.

It is claimed that the corresponding equation \eqref{sDCHE} admits a non-trivial solution holomorphic at zero.

The above   $\lambda,\mu,\omega $ determine the parameters $A,B,\omega$ of the equation \eqref{overdamped RSJ}  in accordance with formulas \eqref{AB vs lamu} (in which $\llll=2$).

It is similarly claimed that all its solutions are periodic \modtwopi{} that corresponds to the point of constriction of phase-lock subarea in the space of equation parameters
distinct from the one we started from although the rotation numbers coincide and are equal to 2.
 In particular, the new frequency differs from (namely, is less than) the original $\omega$.

Another distinction arises due the incrementation  of the constriction level from 1 to 2.

The the above computations constitute a step of the algorithm of constructing of a free-of-bound sequence of constrictions by means of isomonodromic
deformation of special double confluent Heun equation \eqref{sDCHE}. This step can be repeated with initial data ensured by the result of the preceding one.

The result of such a computation is presented in \Fig\ref{fig10}.
The only qualitative distinction is the unboundedness of $\bfuplam(\akat)$ on the left near the pole $\tpole_1$
(in \Fig\ref{fig7}, $ \bfuplam  $ is bounded on the left).
In particular, one more root of $\bfuplam(\akat)$ is again revealed yielding yet another suit of parameters of a constriction
arising
at a less
frequency. The rotation number does not change while, as earlier, the constriction level increases having become now equal to 3.

Although  $\bfuplam$-functions (i.e.\ restrictions of  $\bfuplam(\akat)$ to segments of $\mathbb{R}_+$ bounded by its adjacent poles)
enabling one to discern the location of constrictions
seems to be the most significant outcome of our computation, the role of $\bfupmu$-functions is no less important.

First of all, as opposed to $\bfuplam$ and $\dualam$  harmed by singularities,
these functions
combine to a single continuous and apparently smooth one with values staying inside the interval $[0,1]$.

 The same concerns the `reflected'
$\duamu$-functions also yielding a continuous smooth damping oscillations on the whole available combined domain with values falling in the same segment.
The distinction is the levels near which their graphs  accumulate: this is 1 from  below for $\bfupmu$  and 0 from above for $\duamu$.
Moreover, they osculate these boundary levels that occurs, for $\bfupmu$,  at the poles of $\bfuplam$ and, for  $\duamu$, at the roots.
Equivalent reference is: at the roots  of $\dualam$ for  $\bfupmu$, and at its poles   for $\duamu$.

The noted osculations looks like the images of neighborhoods of second order roots of smooth functions (of the difference $1- \bfupmu$ in the former case).
If this is the case then the half-order powers
$(1-\bfupmu(\akat))^{1/2}$
and
$\duamu(\akat) ^{1/2}$ should exhibit, upon appropriate selection among dual branches, a perfect smoothness.
The conjecture is confirmed by
the corresponding graphs presented in \Fig\ref{fig11}
plotted
employing explicit transformations to recast the already shown images of the same data.

Building on such an apparent success, it is natural to introduce the equations for the above two square roots.
Remarkably,
these equations prove to be coinciding except for the difference in values of the only constant parameter involved therein.
 See definition \eqref{these are bounded} and
\Eq\eqref{the second one}.

Starting from the data corresponding to a root of $\bfuplam(\akat)$ associated with some constriction,
the above equation can be easily integrated
on an interval of variation of $\akat$ of arbitrary `duration'.
The graphs shown in \Fig\ref{fig12} represent such solutions continuing the functions plotted in \Fig\ref{fig11} to a considerably greater domain.
The roots of one of them coincide with roots of $\bfuplam$\;\footnote{and also with roots of $\xi_\llll$}, the roots of another one
specify locations of poles of the same function.
The six roots of $\bfuplam$ and hence locations of the six constrictions has been found. These data are summarized in \Tab\ref{tab2}.
Their further replenishment is feasible.

The integration of pairs of decoupled equations \eqref{the second one} gives rise therefore to an efficient method of realization of
isomonodromic deformation of \Eq\eqref{sDCHE}

\section{Conclusion}

The points of the space of parameters of \Eq\eqref{overdamped RSJ} with given $\omega>0$ situated in constrictions of its phase-lock area
are associated with instances of \Eq\eqref{sDCHE} admitting a non-trivial solution holomorphic at zero and hence characterized by trivial monodromy groups.
Deformations of \Eq\eqref{sDCHE} described by \Eq\eqref{J-Heun deformed} preserve triviality of the latter in case of fulfilment of
\Eqs\eqref{eq for bfuplam}, \eqref{eq for bfupmu}. The decoupled equation for the function-parameter $\bfuplam(\akat)$ represents a particular
instance of \PainIII{} equations.

 These are roots of $ \bfuplam $  at which the equation \eqref{J-Heun deformed} becomes equivalent to equations reducible to \Eq\eqref{sDCHE}
and the latter inherits
the property of triviality of the monodromy group, provided their deformation possesses it.
This means that the instances of RSJ equations \eqref{overdamped RSJ} finally obtained in this way
force all their solutions to be periodic \modtwopi{} meaning that their parameters fall into constrictions of phase-lock areas in their space.
Thus passing through roots of $ \bfuplam $,  an available assembly of the constriction data can be modified to one or more new ones.

In example we have considered \PIIItran{} $ \bfuplam $ has indeed many real positive roots
and it can be conjectured that their full sequence is infinite.
However, there also are simple poles on the positive real half-axis, each of them being situated  between an own pair of adjacent roots.
This feature should be generic as well.

Alternatively, the roots and poles of $ \bfuplam $ can be found from results of integration
of two instances
of the non-linear second order differential equation \eqref{the second one} whose derivation indicates its equivalence
to a subset of the family of \PainIII{} equations.

In our example the solutions to \Eq\eqref{the second one} prove to be bounded on the whole requested domain
and look like damped oscillations around zero level. It can also be conjectured that they are regular at zero argument
(the singular point for the equations they obey) as well.
The roots of one of them coincides with roots of  $ \bfuplam $, the roots of another one are at locations of  $ \bfuplam $ poles.
All the data specifying constrictions can be elicited therefrom.
It particular, the sequence of frequencies thus obtained seems to be worth of a separate study.

\appendix

\section{Examples of typical forms of the flow of integral curves for equation of overdamped Josephson junction}

\begin{figure}[H]
\scalebox{0.85}{
\includegraphics{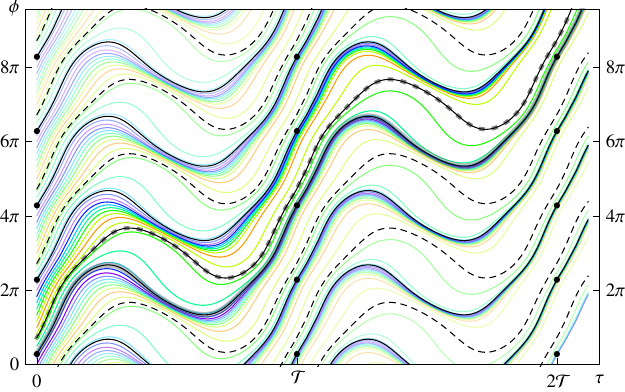}
}
\caption{
\small
A fragment of the flow
formed by integral curves for
\Eq\eqref{sDCHE} in case of $\omega=0.7$, $B=2\omega$, $A=3.66129\!\dots$
The invariance with respect to vertical shift by $2\pi$ takes place.
Each segment on the vertical coordinate axis $\akatau=0$ (not shown) of length $2\pi$
contains 12 uniformly distributed starting points of graphs of computed solutions.
To them, the two additional graphs representing \modtwopi-periodic solutions, stable and unstable, are added.
A stable \modtwopi-periodic solution serving attractor for the other ones starts at $\phi(0)=2\pi\cdot 0.138904\!\dots$
It (as well as similar graphs shifted by integer numbers of $2\pi$) is indicated by points situated at intersections of the graph with verticals $\akatau=0,\Tau,2\Tau$.
The unstable solutions (repellers) are imaged by dashed curves.
One of them starts at $\phi(0)=2\pi\cdot 0.361096\!\dots$\, Other ones can be obtained from it by vertical shifts by multiples of $2\pi$.
The rotation number can be inferred from coinciding increments of \modtwopi-periodic solutions across $\akatau$-lapses $\Tau$. It is equal to 2=$B/\omega$.
}\label{fig14}
\end{figure}

\begin{figure}[h]
\scalebox{0.9}{
\includegraphics{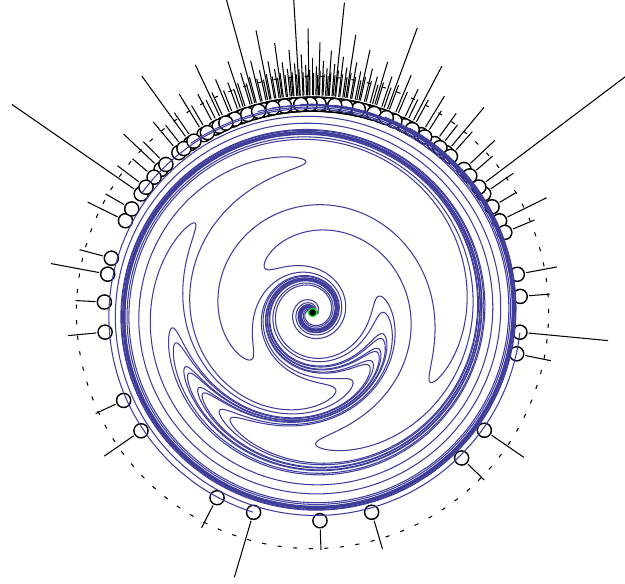}
}
\caption{
\small
The diagram illustrating apparent disorder in development of solutions to \Eq\eqref{sDCHE}
with $\omega=0.7,$ $A=3.85399\!\dots,$  $ B=1.75$ for which
the phase-lock is stably absent.
The solution starting at randomly chosen $\phi(0)=2\pi\cdot  0.34105\!\dots$ and made known for $0\le\akatau\le 128\Tau$ ($\Tau=2\pi/\omega$)
by means of numerical integration
is examined.
Regarding the figure canvas as a part of the complex plane around zero,
12 curves started near the center and ended on the unit circle (not shown)
represent the
graphs of the 
functions $ \delta\akatau/\Tau\cdot e^{\Imi\phi(\delta\akatau + j\Tau)}  $ for $0\le\delta\akatau\le \Tau$ and $j=0,1,\dots,11$
characterizing behavior of $\phi(\akatau)$ on a subsequent pairwise adjacent segments taken from its unbounded domain.
The radial distance encodes here the variation of $\akatau$
on its subsequent `lapses of duration $\Tau$' and the angular interpretation of $\phi$ is explicitly imposed.
The final endpoints of the shown graphs are indicated by small disks centered at them. Similar discs are also put around endpoints of the subsequent graphs
corresponding to $j=12,13,\dots,127$ which are not shown themselves in order to avoid picture cluttering.
Overlapping of discs agrees with positions of the corresponding endpoints in their naturally ordered sequence.
More clear characteristic of their order is the lengths of `the spines' attached radially to all the noted discs from outside which are equal to $(j+1)^{-1/2}$.
The dashed circle indicates the positive minimum of lengths of such `spines' arising in view of finiteness of the number (here 128) of lapses of $\akatau$
taken into account. \\
Discontinuous dependence of the lengths of `spines' on their direction  is the sign of (pseudo-)chaotic behavior of
the solution
$\phi(\akatau)$ under consideration.
}\label{fig15}
\end{figure}

%
\begin{figure}[h]
\scalebox{0.9}{
\includegraphics{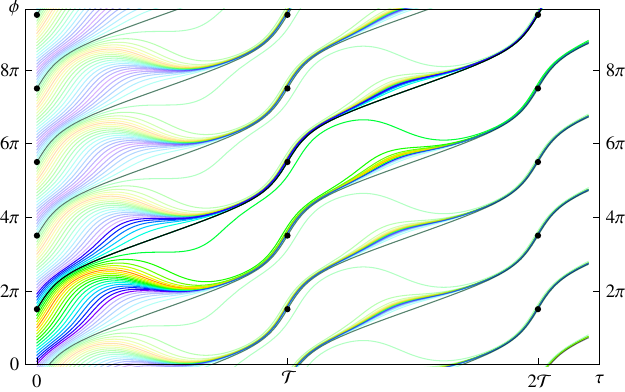}
}
\caption{
\small
A fragment of the flow
formed by integral curves for
\Eq\eqref{sDCHE}
under conditions admitting  solution $\phi=\phi_1$ specified
by formulas \eqref{polycases}.
The parameters are set as follows: $\omega=0.7$, $B=2\omega$, $A=\sqrt{1+2\omega}=1.54919\!\dots$
Upon unbounded continuation,
the invariance with respect to vertical shift by $2\pi$ takes place.
Each segment on the vertical coordinate axis $\akatau=0$ (not shown) of length of $2\pi$
contains 24 uniformly distributed starting points of graphs of solutions.
Extending to the right, the latter come close from above to some graph from the sequence corresponding to \modtwopi-periodic solutions
produced from the noted solution $\phi_1$ by means of vertical $2\pi$-shifts (dots are placed at their intersections with verticals $\akatau=0,\Tau,2\Tau$).
The opposite downsides of graphs
from the mentioned sequence 
repel nearby graphs of other solutions although $\tau$ has to be sufficiently large to make this property vivid.  %
Each vertical segment of length $2\pi$ either have a single point of intersection 
with graph of a \modtwopi-periodic solution or its boundary points are the points of two such intersections.
It holds $\phi(0) \stackrel{\mbox{\tiny \modtwopi}}{ =\; }  3\pi/2$
for the \modtwopi-periodic solutions.
The rate of their growth
characterized by the increments $4\pi$ taking place across $\akatau$-lapses $\Tau$
makes evident the value of the rotation number which is equal to 2.
}\label{fig16}
\end{figure}

\end{document}